\documentclass[twocolumn,showpacs,amsmath,amssymb,amsfonts,nofootinbib,prc]{revtex4-1}
\usepackage{graphicx}% Include figure files
\usepackage{dcolumn}% Align table columns on decimal point
\usepackage{bm}% bold math
\usepackage{amssymb}

\def\beq{\begin{equation}}
\def\eeq{\end{equation}}

\usepackage[usenames]{color}

\newcommand{\be}{\begin{equation}}
\newcommand{\ee}{\end{equation}}
\newcommand{\bea}{\begin{eqnarray}}
\newcommand{\eea}{\end{eqnarray}}
\newcommand{\nn}{\nonumber}

\newcommand \f {\not\!}

 \input{epsf}

\begin{document}
 %%%%%%%%%%%%%%%%%%%%%%%%%%%%%%%%%%%%%%%%%%%%%%%%%%%%%%%%%%%%%%%%%%%%%%%%%%%%%%%%%%%%%%%%%%%%%%%%%%%%
 \title{ Drag induced radiative loss from semi-hard heavy quarks}
 \author{Raktim Abir}
 \author{Abhijit Majumder}
 \affiliation{Department of Physics and Astronomy, Wayne State University, 666 W. Hancock St., Detroit, 
 MI 48201, USA}
 \date{\today}
 %%%%%%%%%%%%%%%%%%%%%%%%%%%%%%%%%%%%%%%%%%%%%%%%%%%%%%%%%%%%%%%%%%%%%%%%%%%%%%%%%%%%%%%%%%%%%%%%%%%%
 \begin{abstract} 
 %
 % \checkmark
 The case of gluon bremsstrahlung off a heavy quark in extended nuclear matter is revisited within 
 the higher twist formalism. In particular, the in-medium modification of ``semi-hard'' heavy quarks 
 is studied, where the momentum of the heavy quark is larger but comparable to the mass of the heavy 
 quark ($p \gtrsim M$). 
 In contrast to all prior calculations, where the gluon emission spectrum is entirely controlled by the  transverse momentum diffusion parameter 
 ($\hat q$), both for light and heavy quarks,  
% The drag coefficient $\hat{e}$ defined as the change in light-cone momentum per unit light-cone path 
% only induces non-radiative loss. 
 %
 in this work, we demonstrate that the gluon emission spectrum for a heavy quark (unlike that for 
 flavors) is also 
 sensitive to $\hat e$, which so far has been used to quantify the amount of \emph{light-cone} drag experienced by a parton. 
 This mass dependent effect, due to the \emph{non-light-like} momentum of a semi-hard heavy-quark, leads to an additional energy loss term for heavy-quarks, while 
 resulting in a negligible modification of light flavor (and high energy heavy flavor) loss. 
 This result can be used to estimate the value of this sub-leading non-perturbative jet transport
 parameter ($\hat e$) from heavy flavor suppression in ultra-relativistic heavy-ion collisions. 
 \end{abstract}
 %%%%%%%%%%%%%%%%%%%%%%%%%%%%%%%%%%%%%%%%%%%%%%%%%%%%%%%%%%%%%%%%%%%%%%%%%%%%%%%%%%%%%%%%%%%%%%%%%%%%
 %%%%%%%%%%%%%%%%%%%%%%%%%%%%%%%%%%%%%%%%%%%%%%%%%%%%%%%%%%%%%%%%%%%%%%%%%%%%%%%%%%%%%%%%%%%%%%%%%%%%
 \maketitle
 %%%%%%%%%%%%%%%%%%%%%%%%%%%%%%%%%%%%%%%%%%%%%%%%%%%%%%%%%%%%%%%%%%%%%%%%%%%%%%%%%%%%%%%%%%%%%%%%%%%%
 %%%%%%%%%%%%%%%%%%%%%%%%%%%%%%%%%%%%%%%%%%%%%%%%%%%%%%%%%%%%%%%%%%%%%%%%%%%%%%%%%%%%%%%%%%%%%%%%%%%%
 \section{Introduction}
 The unprecedented centre of mass energies available at the Large Hadron Collider have opened new 
 windows for the exploration of extreme nuclear matter through high energy jets 
 \cite{Wang:1991xy, Gyulassy:1993hr, Baier:1996kr, Zakharov:1996fv,Majumder:2010qh,Wiedemann:2009sh}. While a large portion of the
 available data on leading (and next-to-leading) particle suppression in the light flavor sector has
 been theoretically described using factorized pQCD based calculations of jet modification 
 \cite{Armesto:2011ht,Qin:2009gw}, 
 heavy quarks have remained somewhat of a challenge \cite{Djordjevic:2013pba}. 
 This is especially true in the semi-hard sector, 
 where the momentum of the heavy 
 quark is larger but comparable to its mass $p \gtrsim m_Q$. 
 We distinguish this region from that of slow heavy quarks, 
 where $p \lesssim m_Q$, 
 which appear to be thermalized with the bulk medium, 
 and fast heavy-quarks, 
 with $p \gg m_Q$ which engender energy loss and suppression similar to light quarks.  
   
 The so called ``heavy-quark puzzle'' had already begun to manifest itself in measurements of 
 the suppression of high transverse momentum (high-$p_T$) non-photonic electrons at the Relativistic 
 Heavy-Ion Collider (RHIC). Measurements by both the STAR \cite{Abelev:2006db} and PHENIX \cite{Adare:2006nq} 
 detectors  showed a slightly higher suppression than 
 expected, based on a calculation that included both drag and radiative loss~\cite{Qin:2009gw,Djordjevic:2003zk,Djordjevic:2004nq}. 
 This trend has continued at the Large Hadron Collider (LHC) where the ALICE experiment has measured $D$ and $B$ meson suppression
 separately, and finds a larger than expected suppression in the semi-hard regime of heavy-quark momentum 
 (we note that the case is not very clear for B-meson suppression which has so far only been presented as 
 $p_{T}$ integrated points)\cite{ALICE:2012ab,Aamodt:2010jd}.

 A considerable amount of theoretical work has been devoted to understand this larger than expected 
 suppression of single electrons or heavy mesons arising from the fragmentation of a heavy-quark. 
 However, most of these may be understood as falling in two categories: 
 Calculations that have extended the base formalism of radiated energy loss for light flavors to include 
 mass dependent terms, as well as a drag term to include the prominent role played by drag in heavy flavor
 energy loss~ ~\cite{Zhang:2004qm,Qin:2009gw,Mustafa:2004dr,Abir:2011jb,Abir:2012pu,Wicks:2005gt}. 
 Calculations that have ignored the role of radiative loss and only focussed on drag loss~\cite{He:2012xz,Cao:2011et,Moore:2004tg}. 
 %[REF papers by Bass, Rapp, Teaney]  {need ref of the paper}
 
 In all calculations above, radiative loss is stimulated by transverse momentum diffusion experienced by the heavy quark or radiated gluon, which, in some cases, is quantified by the jet transport coefficient $\hat{q}$~\cite{Baier:2002tc,Majumder:2012sh}. The drag loss is quantified using the drag coefficient referred to as $dE/dx$ (energy loss per unit distance) or $\hat{e}$~\cite{Majumder:2008zg}.

 To the best of our knowledge, 
 no calculation of heavy flavor energy loss has explored the possibility that the drag coefficient $\hat{e}$ 
 (or the longitudinal diffusion coefficient $\hat{e}_2$) 
 may lead to an additional source of radiative loss, beyond that provided by $\hat{q}$. 
 This possibility is immediately clear in the higher twist framework, 
 where the drag (and longitudinal diffusion) coefficient $\hat{e}$ ($\hat{e}_2$) has the boost invariant definition 
 as the loss of light-cone momentum (fluctuation in light-cone momentum) per unit light-cone length, (assuming a parton moving in the negative light-cone direction)
 \bea
 \hat{e} = \frac{d  \langle \Delta p^- \rangle }{dL^-},  \,\,\,\,\,\,\,\,\,\,  \hat{e}_2 = \frac{d \langle \Delta {p^-}^2 \rangle}{dL^-} .
 \eea
 While such a transport coefficient leads to little change in the off-shellness of a near on-shell \emph{massless} quark, it has a considerable impact on the off-shellness of a near on-shell \emph{massive} quark. No doubt, such a term will \emph{only} have an effect on the radiative 
 loss of a patron where the momentum $p$ is comparable to the mass $M$, thus for light flavors, and for energetic heavy-quarks where, 
 $p \gg M$, this term will have a minimal effect. This was explicitly explored for photon radiation from a light quark in Ref.~\cite{Qin:2014mya}. 
 We point out that such an effect is by no means limited to the higher-twist scheme, but effects several other formalisms that have considered the radiative 
 loss from a heavy-quark in a quark gluon plasma.

 In this paper, 
 we will explore the modification to the calculation of radiative loss, due to the presence of this additional source 
 within the higher twist formalism. 
  To delineate  the importance of these terms, we will use power-counting techniques borrowed from Soft-Collinear-Effective-Theory (SCET)~\cite{Bauer:2000yr,Bauer:2001ct,Bauer:2002nz,Bauer:2001yt} 
  to identify the regime where these mass dependent terms 
will cause detectable effects on the gluon bremsstrahlung spectrum. 
 As a first attempt, we will consider only the case of single scattering and single emission. 
 Arguments presented in subsequent sections will demonstrate this to be the leading contribution, 
 given the short formation time of the radiated gluons.
 In this paper, the analytical expressions for the $\hat{e}$ ($\hat{q}$ and $\hat{e}_2$) induced gluon radiation 
 spectrum will be derived. 
 Numerical calculations for the suppression of  $B$ and $D$ mesons, as well as the suppression and azimuthal anisotropy of 
 non-photonic leptons from the decay of these mesons at LHC and RHIC energies, and comparisons with experimental data will be carried out in a subsequent effort. 

% %
 The article is organized as follows: 
 in Sec.~II, we will setup the basic formalism of Deep Inelastic Scattering (DIS) on a large nucleus, where the hard virtual photon 
 strikes a heavy-quark, assumed to be produced in a rare high $Q^{2}$ fluctuation inside a proton. 
 In Sec.~III, we will carry out diagrammatic studies on the induced gluon radiation off the heavy quark in this 
 system, within the higher twist formalism. 
 We will present and discuss final expressions for gluon bremsstrahlung from such a heavy quark, containing both $\hat q$, 
 $\hat e$ and $\hat{e}_2$ in Sec.~IV. 
 We offer concluding discussions and an outlook in Sec.~V.
 Involved expressions for a set of diagrams are contained in the appendix. 
 
 \section{Deep inelastic scattering and the semi-hard heavy quark}
 %%%%%%%%%%%%%%%%%%%%%%%%%%%%%%%%%%%%%%%%%%%%%%%%%%%%%%%%%%%%%%%%%%%%%%%%%%%%%%%%%%%%%%%%%%%%%%%%%%%%
 % \checkmark 
 The set-up is based on the deep-inelastic scattering of a virtual photon off a heavy quark within a 
 large nucleus with mass number $A$.
 %
 % \checkmark %%%%%%%%%%%%%%%%%%%%%%%%%%%%%%%%%%%%%%%%%%%%%%%%%%%%%%%%%%%%%%%%%%%%%%%%%%%%%%%%%%%%%%%
 We will study the case where the hard virtual photon scatters with the hard heavy quark converting it to 
 slow moving heavy quark (this is defined below).
 %
 % \checkmark %%%%%%%%%%%%%%%%%%%%%%%%%%%%%%%%%%%%%%%%%%%%%%%%%%%%%%%%%%%%%%%%%%%%%%%%%%%%%%%%%%%%%%%
 The propagation of the heavy-quark will be factorized from the hard 
 scattering vertex which produces the outgoing slow moving heavy-quark. 
 %
 % \checkmark %%%%%%%%%%%%%%%%%%%%%%%%%%%%%%%%%%%%%%%%%%%%%%%%%%%%%%%%%%%%%%%%%%%%%%%%%%%%%%%%%%%%%%%
 The nucleus has a momentum $P = pA$, where $p$ is the average momentum of a nucleon in this nucleus.
 %
 % \checkmark %%%%%%%%%%%%%%%%%%%%%%%%%%%%%%%%%%%%%%%%%%%%%%%%%%%%%%%%%%%%%%%%%%%%%%%%%%%%%%%%%%%%%%%
 In the Breit frame, the exchanged virtual photon possesses no transverse momentum,
 %
 % \checkmark %%%%%%%%%%%%%%%%%%%%%%%%%%%%%%%%%%%%%%%%%%%%%%%%%%%%%%%%%%%%%%%%%%%%%%%%%%%%%%%%%%%%%%%
 \begin{eqnarray}
 q \equiv [q^{+}, q^{-}, q_{\perp}] = \left[q^{+},q^{-},0\right].
 \end{eqnarray} 
 %
 % \checkmark %%%%%%%%%%%%%%%%%%%%%%%%%%%%%%%%%%%%%%%%%%%%%%%%%%%%%%%%%%%%%%%%%%%%%%%%%%%%%%%%%%%%%%%
 The process under consideration is the following: 
 %
 % \checkmark %%%%%%%%%%%%%%%%%%%%%%%%%%%%%%%%%%%%%%%%%%%%%%%%%%%%%%%%%%%%%%%%%%%%%%%%%%%%%%%%%%%%%%%
 \begin{eqnarray}
 e(L_1) + A(P) \rightarrow e(L_2) + J_{\cal Q}(L_{\cal Q}) + X .
 \label{chemical_eqn}
 \end{eqnarray}
 %
 % \checkmark %%%%%%%%%%%%%%%%%%%%%%%%%%%%%%%%%%%%%%%%%%%%%%%%%%%%%%%%%%%%%%%%%%%%%%%%%%%%%%%%%%%%%%%
 In the above process, $e(L_{1})$ and $e(L_{2})$ represent the incoming (outgoing) electron with 
 momentum $L_{1}$ and $L_{2}$ respectively. 
 %
 % \checkmark %%%%%%%%%%%%%%%%%%%%%%%%%%%%%%%%%%%%%%%%%%%%%%%%%%%%%%%%%%%%%%%%%%%%%%%%%%%%%%%%%%%%%%%
The factor $A(P)$ represents the incoming nucleus with momentum $P$. 
 %
 % \checkmark %%%%%%%%%%%%%%%%%%%%%%%%%%%%%%%%%%%%%%%%%%%%%%%%%%%%%%%%%%%%%%%%%%%%%%%%%%%%%%%%%%%%%%%
The factor $J_{\cal Q} (L_{\cal Q})$ is the outgoing jet which contains one heavy quark $\cal Q$. In the particular diagrams that we 
will consider, the final strongly interacting particles    
 %
 % \checkmark %%%%%%%%%%%%%%%%%%%%%%%%%%%%%%%%%%%%%%%%%%%%%%%%%%%%%%%%%%%%%%%%%%%%%%%%%%%%%%%%%%%%%%%
% To produce a jet containing a single heavy-quark, there are no available valence heavy-quarks within 
% the nucleus. 
 %
 % \checkmark %%%%%%%%%%%%%%%%%%%%%%%%%%%%%%%%%%%%%%%%%%%%%%%%%%%%%%%%%%%%%%%%%%%%%%%%%%%%%%%%%%%%%%%
 Due to the absence of valence heavy-quarks within the nucleon, the photon will have to strike a heavy quark from s ${\cal Q} \bar{\cal Q}$ fluctuation within the sea of partons. Alternatively one may consider the case of a $J/\psi$ or $\Upsilon$ state bound within a large nucleus, being struck by the hard virtual photon.
 %
 % \checkmark %%%%%%%%%%%%%%%%%%%%%%%%%%%%%%%%%%%%%%%%%%%%%%%%%%%%%%%%%%%%%%%%%%%%%%%%%%%%%%%%%%%%%%%
 More detail discussions on the production of heavy quark can be found in \cite{Abir:2014sxa}. 
 % 
 % \checkmark %%%%%%%%%%%%%%%%%%%%%%%%%%%%%%%%%%%%%%%%%%%%%%%%%%%%%%%%%%%%%%%%%%%%%%%%%%%%%%%%%%%%%%%
 In this work, we will not discuss the production of the heavy quark further. 
 For the purposes of this calculation, this is now contained within a parton distribution function. 
 In essence, by this mechanism a semi-hard heavy quark has been produced. 
  % \checkmark %%%%%%%%%%%%%%%%%%%%%%%%%%%%%%%%%%%%%%%%%%%%%%%%%%%%%%%%%%%%%%%%%%%%%%%%%%%%%%%%%%%%%%%
 In what follows, we will focus on the power counting of the momentum components and the modification of the final state.  
 
 % \checkmark %%%%%%%%%%%%%%%%%%%%%%%%%%%%%%%%%%%%%%%%%%%%%%%%%%%%%%%%%%%%%%%%%%%%%%%%%%%%%%%%%%%%%%%
Similar to Ref~\cite{Abir:2014sxa}, we will consider a quark mass $M \gg \Lambda_{QCD}$ and a final 
 outgoing quark momentum which is larger, but of the order of the quark mass. 
 %
 % \checkmark %%%%%%%%%%%%%%%%%%%%%%%%%%%%%%%%%%%%%%%%%%%%%%%%%%%%%%%%%%%%%%%%%%%%%%%%%%%%%%%%%%%%%%%
 We assume that the quark, anti-quark fluctuations possess minimal momentum in the rest frame of 
 the nucleus, and thus its momentum components scales as $\left(p_{\cal Q}^{+},p_{\cal Q}^{-},
 p_{\cal Q \perp} \right)\sim \left(M/\sqrt{2},~M/\sqrt{2},~0\right)$.
 %
 % \checkmark %%%%%%%%%%%%%%%%%%%%%%%%%%%%%%%%%%%%%%%%%%%%%%%%%%%%%%%%%%%%%%%%%%%%%%%%%%%%%%%%%%%%%%%
 Now in a frame where the nucleus is boosted by a large boost factor $\gamma$ in the 
 $``+\textquotedblright$ direction, momentum components of the incoming heavy quark will scales as, 
 %
 % \checkmark %%%%%%%%%%%%%%%%%%%%%%%%%%%%%%%%%%%%%%%%%%%%%%%%%%%%%%%%%%%%%%%%%%%%%%%%%%%%%%%%%%%%%%%
 \begin{eqnarray}
 p_{\cal Q} = \left[ p_{\cal Q}^{+}, p_{\cal Q}^{-}, p_{\cal Q \perp} \right] \equiv
 \left[ \gamma \frac{M}{\sqrt{2}},~\frac{1}{\gamma}\frac{M}{\sqrt{2}},~0   \right].
 \end{eqnarray}
 %
 % \checkmark %%%%%%%%%%%%%%%%%%%%%%%%%%%%%%%%%%%%%%%%%%%%%%%%%%%%%%%%%%%%%%%%%%%%%%%%%%%%%%%%%%%%%%%
 It is important to note that the boost factor $\gamma$ carries no extra information other than 
 the the fact that $p_{\cal Q}^{+}$ is very large compare to $p_{\cal Q}^{-}$.
 %
 % \checkmark %%%%%%%%%%%%%%%%%%%%%%%%%%%%%%%%%%%%%%%%%%%%%%%%%%%%%%%%%%%%%%%%%%%%%%%%%%%%%%%%%%%%%%%
 We assume the momentum components of the incoming photon to be,
 %
 % \checkmark %%%%%%%%%%%%%%%%%%%%%%%%%%%%%%%%%%%%%%%%%%%%%%%%%%%%%%%%%%%%%%%%%%%%%%%%%%%%%%%%%%%%%%%
 \begin{eqnarray}
 q = \left[ -\gamma \frac{M}{\sqrt{2}} + \frac{M^{2}}{2q^{-}},
 ~ q^{-} - \frac{1}{\gamma}\frac{M}{\sqrt{2}}, ~0   \right] .
 \end{eqnarray}
 %
 % \checkmark %%%%%%%%%%%%%%%%%%%%%%%%%%%%%%%%%%%%%%%%%%%%%%%%%%%%%%%%%%%%%%%%%%%%%%%%%%%%%%%%%%%%%%%
Given a large boost factor $(\gamma)$, one can assume that $\gamma M \gg M \sim q^{-} \gg 
 M/\gamma $. 
 %
 % \checkmark %%%%%%%%%%%%%%%%%%%%%%%%%%%%%%%%%%%%%%%%%%%%%%%%%%%%%%%%%%%%%%%%%%%%%%%%%%%%%%%%%%%%%%%
 Hence, we define the hard scale $Q$ as  $Q^2 = - q^2 \simeq \gamma M q^{-}/\sqrt{2}$. As a result, 
 we obtain the final out-going quark to have momentum components as,  
 \begin{eqnarray}
 p_f=p_{\cal Q} + q = \left[\frac{M^{2}}{2q^{-}} , q^{-} , 0 \right].
 \end{eqnarray}
 %
 % \checkmark %%%%%%%%%%%%%%%%%%%%%%%%%%%%%%%%%%%%%%%%%%%%%%%%%%%%%%%%%%%%%%%%%%%%%%%%%%%%%%%%%%%%%%%
 We consider $M \lesssim q^{-} \sim \sqrt{\lambda} Q$ where $\lambda \sim 1/\gamma$ for this 
``semi-hard" heavy quark. The term ``semi-hard'' defines a quark whose momentum is not an order of 
 magnitude larger than its mass. 

 %%%%%%%%%%%%%%%%%%%%%%%%%%%%%%%%%%%%%%%%%%%%%%%%%%%%%%%%%%%%%%%%%%%%%%%%%%%%%%%%%%%%%%%%%%%%%%%%%%%%
 %%%%%%%%%%%%%%%%%%%%%%%%%%%%%%%%%%%%%%%%%%%%%%%%%%%%%%%%%%%%%%%%%%%%%%%%%%%%%%%%%%%%%%%%%%%%%%%%%%%%
 \vspace{0.5cm}
 \subsection{Power Counting and the small $\lambda$ parameter} 
 %
 % \checkmark %%%%%%%%%%%%%%%%%%%%%%%%%%%%%%%%%%%%%%%%%%%%%%%%%%%%%%%%%%%%%%%%%%%%%%%%%%%%%%%%%%%%%%%
 In order to set up the power counting, in this study, we have introduced the dimensionless small
 parameter $\lambda$. 
 %
 % \checkmark %%%%%%%%%%%%%%%%%%%%%%%%%%%%%%%%%%%%%%%%%%%%%%%%%%%%%%%%%%%%%%%%%%%%%%%%%%%%%%%%%%%%%%%
 Power corrections to hard process are generally suppressed by factors of a hard scale,
 $Q^2 \gg \Lambda_{QCD}$.
 The introduction of the parameter $\lambda$ to represent semi-hard scales as $\lambda Q$ and softer 
 scales as $\lambda^2 Q$, is a concept borrowed from soft collinear effective theory (SCET)~
 \cite{Idilbi:2008vm,D'Eramo:2010ak}.
 In what follows, we will retain leading and next-to-leading terms in $\lambda$ power counting, neglecting all terms which 
 scale with  $\lambda^2$ or a higher power of $\lambda$~\cite{Qin:2012fua}.
 %
 % \checkmark %%%%%%%%%%%%%%%%%%%%%%%%%%%%%%%%%%%%%%%%%%%%%%%%%%%%%%%%%%%%%%%%%%%%%%%%%%%%%%%%%%%%%%%
%
% \item Terms that are sub-leading in two order of magnitude can be dropped but the terms that are 
% suppressed by one oder of magnitude should be kept. 
% \end{itemize}
 %
 % \checkmark %%%%%%%%%%%%%%%%%%%%%%%%%%%%%%%%%%%%%%%%%%%%%%%%%%%%%%%%%%%%%%%%%%%%%%%%%%%%%%%%%%%%%%%
 We have chosen the scaling variable $\lambda$ in such a way that perturbation theory may be applied 
 down to momentum transfer scales at or above $\lambda^{3/2}Q\sim \Lambda_{\rm QCD}$.
 In our study,
 % \checkmark %%%%%%%%%%%%%%%%%%%%%%%%%%%%%%%%%%%%%%%%%%%%%%%%%%%%%%%%%%%%%%%%%%%%%%%%%%%%%%%%%%%%%%%
 \begin{eqnarray}
 \lambda^0 \gg \sqrt{\lambda} \gg \lambda \gg {\lambda}^{\frac{3}{2}}  .
 \end{eqnarray}
 %
 % \checkmark %%%%%%%%%%%%%%%%%%%%%%%%%%%%%%%%%%%%%%%%%%%%%%%%%%%%%%%%%%%%%%%%%%%%%%%%%%%%%%%%%%%%%%%

 Based on the power counting set up and the choice of incoming and outgoing quark momentum, 
 we can outline several important scales in the problem of a heavy quark propagating through the 
 nuclear medium, scattering off constituents in the medium and 
 emitting real gluons.
 %%%%%%%%%%%%%%%%%%%%%%%%%%%%%%%%%%%%%%%%%%%%%%%%%%%%%%%%%%%%%%%%%%%%%%%%%%%%%%%%%%%%%%%
 We remind the reader that given the semi-hard momentum of the heavy quark, collinear emission is 
 suppressed due to the mass of the heavy quark, i.e., the \emph{dead cone} effect~\cite{Dokshitzer:2001zm}.
 %
 %%%%%%%%%%%%%%%%%%%%%%%%%%%%%%%%%%%%%%%%%%%%%%%%%%%%%%%%%%%%%%%%%%%%%%%%%%%%%%%%%%%%%%%%
 In the subsequent section, the calculation of the process of single scattering and single emission from a heavy quark will be carried out. 
 Here we outline the power counting (in $\lambda$) of the relevant momentum components that will arise in the calculation.
 The virtuality of the hard photon defines the hardest scale in the problem, $Q$, similar to the case of light quark production in DIS. 
 The incoming or initial heavy quark has momentum components $p_i\sim(\lambda^{-\frac{1}{2}},\lambda^{\frac{3}{2}},0)Q$, the outgoing heavy quark has momentum components $p\sim(\sqrt{\lambda},\sqrt{\lambda},0)Q$. 
 The mass of the semi-hard heavy quark scales as $M\sim \sqrt{\lambda} Q$. 
 This choice of incoming parton and photon momenta ensures that the momentum of the final outgoing heavy quark is of the order of its momentum. 
 In what follows, we will demonstrate that the leading contribution to gluon emission arises from the 
 region where real emitted gluons have momenta which scale as $l \sim(\lambda,\lambda,\lambda)Q$.
 The fraction of light cone momenta carried out by the gluon is $y=l^-/p^- \sim \sqrt{\lambda}$.
 Also somewhat different from the case of light flavors is the scaling of the virtual Glauber gluons : $k\sim(\lambda^{\frac{3}{2}},\lambda^{\frac{3}{2}},\lambda)Q$  with $k^2=2k^+k^--k_\perp^2 \simeq -k_\perp^2$.  
 As we will demonstrate below, these choices of momentum scales tend to enhance the gluon emission rate.

There is another consequence of this choice of scales which relates to how the single gluon emission kernel may be iterated. 
Unlike the case for light flavors, the formation length of a gluon with momentum components $l \sim (\lambda, \lambda, \lambda) Q$ is
\bea
\tau_Q = \frac{2 l^-}{l_\perp^2} \sim \frac{1}{\lambda Q} ,
\eea
is rather short compared to the formation length of $\tau_q\sim~1/\lambda^2 Q$ for gluon radiation from near on-shell light flavors. 
This indicates that there cannot be many scatterings per emission. As a result, in what follows, we derive the single scattering per gluon emission rate. This single gluon emission kernel, induced by single scattering will have to be iterated to obtain the full energy loss of a 
semi-hard quark.

 %
 % \checkmark %%%%%%%%%%%%%%%%%%%%%%%%%%%%%%%%%%%%%%%%%%%%%%%%%%%%%%%%%%%%%%%%%%%%%%%%%%%%%%%%%%%%%%%
 Many readers may find the presence of factors of $\sqrt{\lambda}$ somewhat disconcerting. 
 We could have simply replaced this with a new $\lambda$. 
 We refrain from defining a new dimensionless parameter $\lambda$, so as to make contact with prior definitions of 
  $\lambda$ used in the case of light quarks, where $\lambda Q $ represented the transverse momentum of the 
  radiated gluons from a hard parton, or the transverse momentum from scattering of a gluon in the medium. Thus, 
  to continue to draw a parallel with the prior results form light flavor energy loss, we require $\lambda Q \sim 1$~GeV.

 To get a physical feel of these scaling relations, one may typically assume
 $Q \sim 100~{\rm GeV},~\sqrt{\lambda} Q\sim 10~{\rm GeV},~\lambda Q\sim 1~{\rm GeV},~\lambda^{\frac{3}{2}} Q\sim {\rm \Lambda_{QCD}} $.
 %
 % \checkmark %%%%%%%%%%%%%%%%%%%%%%%%%%%%%%%%%%%%%%%%%%%%%%%%%%%%%%%%%%%%%%%%%%%%%%%%%%%%%%%%%%%%%%%
 We will retain terms that are ${\cal O}(\sqrt{\lambda})$ suppressed compared to the leading terms, 
 but ignore all terms that are suppressed by ${\cal O}(\lambda)$ and higher. 
 %
 % \checkmark %%%%%%%%%%%%%%%%%%%%%%%%%%%%%%%%%%%%%%%%%%%%%%%%%%%%%%%%%%%%%%%%%%%%%%%%%%%%%%%%%%%%%%% 
 Thus, terms with $M^2/Q^2 \sim \lambda$ will eventually be ignored.

 \section{Induced Gluon radiation off the heavy quark}
 % \checkmark %%%%%%%%%%%%%%%%%%%%%%%%%%%%%%%%%%%%%%%%%%%%%%%%%%%%%%%%%%%%%%%%%%%%%%%%%%%%%%%%%%%%%%%
 In this section we will discuss some contributions to the next-to-leading order
correction to semi-inclusive 
DIS on a large nucleus with a quark and gluon in the final state. By next-to-leading order, we simply mean 
including one interaction term in the amplitude and complex conjugate, which converts a  
single quark to a quark and a gluon. 
The diagrams we will consider will also contain two scatterings for the full cross section, with 
both scatterings in the amplitude, both in the complex conjugate, or one in amplitude and one in complex conjugate. 
The double differential cross section for the semi inclusive process of an electron with in-coming momentum $L_1$ and out-going momentum of $L_2$, undergoing 
DIS off a nucleus (with momentum $pA$), 
leading to the production of a final state heavy quark with transverse momentum $l_{Q_\perp}$  and a final 
state gluon with transverse momentum $l_\perp \gg \Lambda_{QCD}$, may be expressed as 

\bea
\frac{E_{L_2} d \sigma } {d^3 L_2 d^2 l_{Q_\perp}  d^2 l_\perp dy  } &=&
\frac{\alpha_{em} ^2}{2\pi s  Q^4}  L_{\mu \nu}  
\frac{d W^{\mu \nu}}{d^2 l_{Q_\perp} d^2 l_\perp dy}. \label{LO_cross}
\eea

\noindent
In the equation above, $s = (p+L_1)^2$ is the total invariant mass of the lepton nucleon 
system. In the single photon exchange approximation, the leptonic part of the cross section is 
easily expressed in terms of the leptonic tensor denoted as $L^{\mu \nu}$, given as,
\bea 
L_{\mu \nu} = \frac{1}{2} {\rm Tr} [ \f L_1 \gamma_{\mu} \f L_2 \gamma_{\nu}].
\eea
 \begin{figure}[ht]
   \begin{center}
 \includegraphics[width=8cm,height=4cm]{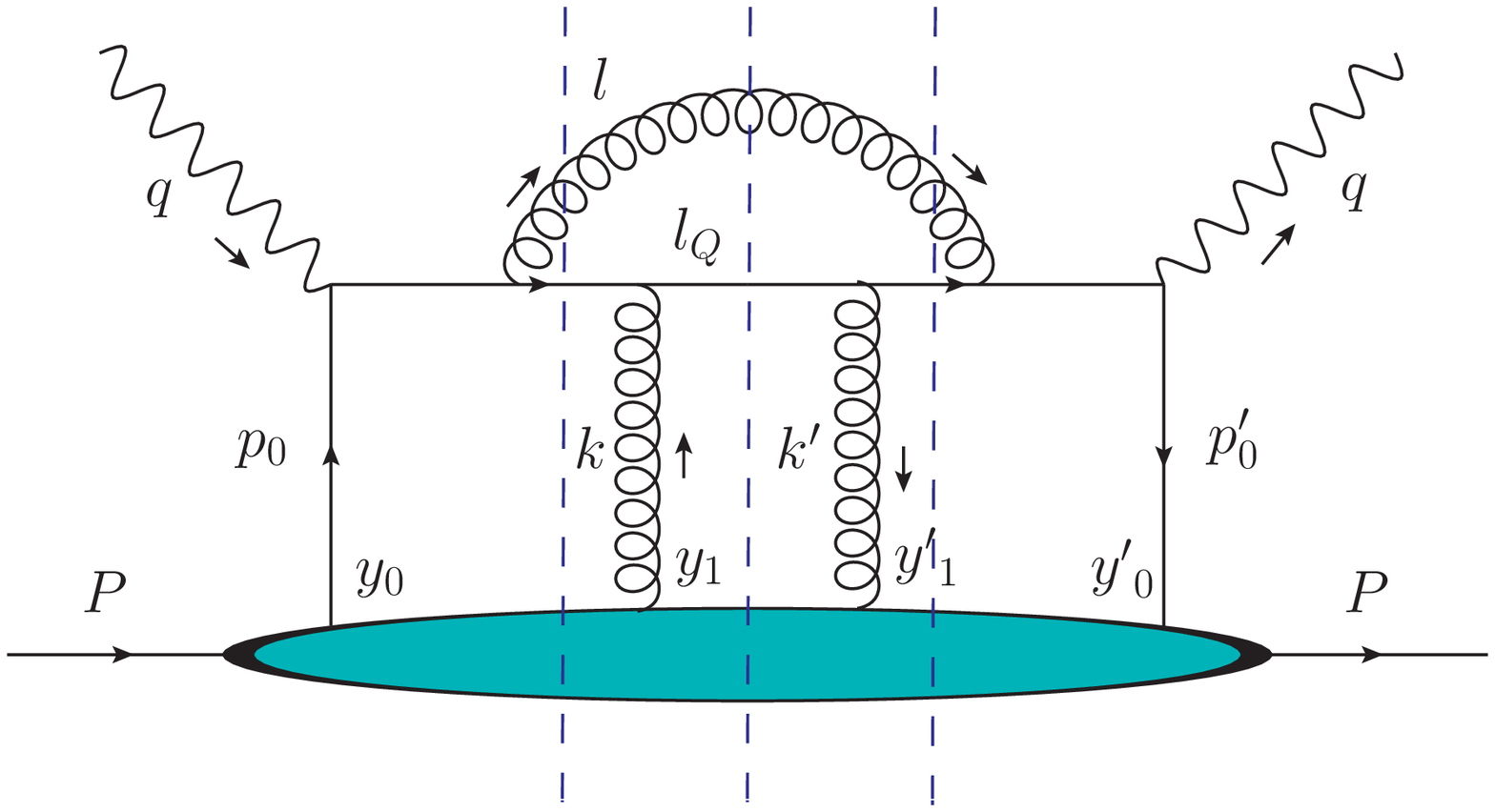}
 \caption{A representative single gluon emission diagram, where gluon emission is induced by single scattering.This represents a symmetric diagram with scattering off the final produced quark. Three separate cuts, denoted as central, left and right are indicated by the dashed lines.}
      \label{Sample_Diagram}
  \end{center}
 \end{figure}

 In what follows, the focus will lie entirely on the hadronic tensor $W^{\mu \nu}$. We will carryout calculations of 
 a set of contributions to $W^{\mu \nu}$ at next-to-leading order (NLO) as described above, and next-to-leading twist (NLT), meaning double scattering in the cross-section.
 
 Already, at NLO and NLT, there are several interfering diagrams to consider. 
 In this section, the calculation of one of the diagrams that contribute to single scattering induced single gluon emission will be 
 carried out in some detail to familiarize the reader to the approximations carried out in this article. 
 Figure~(\ref{Sample_Diagram}) represents the diagram that will be evaluated. 
 This diagram corresponds to the process where a semi-hard heavy quark produced after DIS radiates a 
 gluon followed by a scattering in the medium, and finally exits the nucleus. 
 
 In this study, calculations will be carried out in axial gauge, $n \cdot A=0$, with $n\equiv(1,0,0_\perp)$ and 
 $A^{-}=0$. 
  In $A^-=0$ gauge, the double scattering of a quark radiating a gluon contains a total of nine central cut diagrams where the cut line 
 lies between the two scatterings. It also contains seven left cut diagrams and seven right cut diagrams. 
 In this section, the central cut diagram of Fig.[\ref{Sample_Diagram}] will be analyzed in detail.
  In this section and what follows, only the real contribution where the radiated gluon line has been cut will be considered. The entire contribution from virtual diagrams, which contain a quark gluon fluctuation either in the amplitude or complex conjugate, will be obtained using unitarity arguments. 
 
 This NLO-NLT contribution to the hadronic tensor may be expressed as, 
 \begin{widetext}
 \begin{eqnarray}
 W^{\mu\nu} &=& g^4(-g_{\perp}^{\mu \nu}) \int \frac{d^4{y'_0}d^4p_i'}{(2\pi)^4} \frac{d^4{y'_1}d^4k'}{(2\pi)^4}  
 \frac{d^4{y_1}d^4k}{(2\pi)^4}  \frac{d^4{y_0}d^4p_i}{(2\pi)^4} 
 \frac{d^4l}{(2\pi)^4}\frac{d^4l_Q}{(2\pi)^4} e^{ip'_iy'_0}e^{ik'y'_1}e^{-iky_1} e^{-ip_iy_0}   
 \nn \\
 && \times {\rm Tr} \left[\frac{1}{2}\gamma^- \frac{-i({p \!\!\! /}_f+M )}{p_f^2-M^2-i\epsilon}
    \gamma_{\alpha} \frac{-i({p \!\!\! /}_f-{l \!\!\! /}+M )}{(p_f-l)^2-M^2-i\epsilon} 
   {l \!\!\! /}_q G^{\alpha\beta}
  \frac{i({p \!\!\! /}_f-{l \!\!\! /}+M )}{(p_f-l)^2-M^2+i\epsilon} \gamma_{\beta}
    \frac{i({p \!\!\! /}_f-{l \!\!\! /}+M )}{(p_f-l)^2-M^2+i\epsilon} \right]  \nn \\
  && \times  2\pi \delta(l^2) 2\pi \delta(l_Q^2-M^2) (-i T^a)  (i T^a) 
 \langle A | {\bar \psi}(y_0') \gamma^+ A^{+}(y_1')  A^{+}(y_1) \psi(y_0) | A \rangle.
 \end{eqnarray}
 \end{widetext}
 
 % \checkmark %%%%%%%%%%%%%%%%%%%%%%%%%%%%%%%%%%%%%%%%%%%%%%%%%%%%%%%%%%%%%%%%%%%%%%%%%%%%%%%%%%%%%%%
 We have defined the following momentum fractions for convenience, 
 %
 % \checkmark %%%%%%%%%%%%%%%%%%%%%%%%%%%%%%%%%%%%%%%%%%%%%%%%%%%%%%%%%%%%%%%%%%%%%%%%%%%%%%%%%%%%%%%
 \begin{eqnarray}
 y &=& \frac{l^-}{q^-}~,~~
 \eta = \frac{k^-}{l^-} ~,~~   \zeta = \frac{1-y}{1-y+\eta y}~, \label{momentum-fractions} \\
 x_0 &=& \frac{p_i^+}{P^+}~,~~
 x_1 = \frac{k^+}{P^+}~,~~   \chi=\frac{y^2M^2}{l_\perp^2}~, \nn \\
 x_L &=& \frac{l_\perp^2}{2P^+q^-y(1-y)} ~,~~
 x_D = \frac{k_\perp^2-2l_\perp k_\perp}{2P^+q^-}  ~~,~ \nn \\
 x_K &=& \frac{k_\perp^2}{2P^+q^-} ~,~~\kappa=\frac{1}{1+(1-y)^2}~,~~
 x_M = \frac{M^2}{2P^+q^-}. \nn
 \end{eqnarray}

In what follows, we will simply both numerator and denominator by retaining only leading 
terms in the $\lambda$ power counting highlighted above. This is followed by taking the 
trace in the numerator and contour integrations to simplify the denominator. 
 
The factor $G^{\alpha \beta}$ is the polarization tensor, and in $A^-=~\!\!0$ gauge, it has the form, 
 % \checkmark %%%%%%%%%%%%%%%%%%%%%%%%%%%%%%%%%%%%%%%%%%%%%%%%%%%%%%%%%%%%%%%%%%%%%%%%%%%%%%%%%%%%%%%
 \begin{eqnarray}
 G^{\alpha\beta}=-g^{\alpha\beta}+\frac{n^\alpha l^\beta + n^\beta l^\alpha}{nl}.
 \end{eqnarray}
 %
 % \checkmark %%%%%%%%%%%%%%%%%%%%%%%%%%%%%%%%%%%%%%%%%%%%%%%%%%%%%%%%%%%%%%%%%%%%%%%%%%%%%%%%%%%%%%%
The structure of $n\equiv(1,0,0_\perp)$ implies $n \cdot l=l^-$. Within this kinematic set up, in $A^-=0$ gauge, the leading 
 component of the gluon field is $A^+$. 
 
 % \checkmark %%%%%%%%%%%%%%%%%%%%%%%%%%%%%%%%%%%%%%%%%%%%%%%%%%%%%%%%%%%%%%%%%%%%%%%%%%%%%%%%%%%%%%%
 The spin sum in the numerator, containing the entire set of $\gamma$ matrices from the quark propagators, together with
 those from the interaction with the gauge field may be partially simplified and expressed as, 
 \begin{widetext}
 \begin{eqnarray}
 {{\cal N}_{11}^{c}}&=&{\rm Tr}\left[\frac{1}{2}\gamma^-\left\{\gamma^+q^-\gamma_{\mu}\frac{n^\mu (l_\perp)^\nu}{l^-}
             \gamma^+\left(q^--l^-\right)+\gamma_\perp^\alpha {{p_f}_\perp}_\alpha
              {\gamma_\perp}_\mu \left(-g_\perp^{\mu\nu}\right)\gamma^+\left(q^--l^-\right)  \right. \right. \nn \\
 && \left. \left. +       \gamma^+q^-  {\gamma_\perp}_\mu \left(-g_\perp^{\mu\nu}\right) \gamma_\perp^{\alpha}
 \left({p_f}_\perp-l_\perp\right)_\alpha\right\} \gamma^- \gamma^+(q^--l^-+k^-)\gamma^- 
                  \left\{ \gamma^+(q^--l^-)\gamma_\rho \frac{n^\rho(l_\perp)_\nu}{l^-} \gamma^+q^-  \right.    \right.    \nn \\
 && \left. \left.  +  \gamma^+ (q^--l^-)\gamma_\perp^{\rho}(-g_\perp)_{\rho \nu} \gamma_{\perp}^{\beta}{{p_f}_\perp}_\beta
            +\gamma_\perp^{\beta}\left({p_f}_\perp-l_\perp\right)_\beta \gamma_\perp^{\rho}(-g_\perp)_{\rho\nu} \gamma^+ 
            q^-\right\}\right] \nn \\
 &~& ~ + {\rm Tr}\left[\frac{1}{2}\gamma^-\left\{M {\gamma_\perp}_\mu \left(-g_\perp^{\mu\nu}\right)\gamma^+\left(q^--l^-\right)
                                                +       \gamma^+q^-  {\gamma_\perp}_\mu \left(-g_\perp^{\mu\nu}\right) M   \right\} \right. \nn \\
 && \left.~ \times \gamma^- \gamma^+(q^--l^-+k^-)\gamma^- 
                  \left\{   
    \gamma^+ (q^--l^-)\gamma_\perp^{\rho}(-g_\perp)_{\rho \nu} M
            +M \gamma_\perp^{\rho}(-g_\perp)_{\rho\nu} \gamma^+ 
            q^-\right\}\right] .
 \label{nu11}
 \end{eqnarray}
 \end{widetext}
 
 % \checkmark %%%%%%%%%%%%%%%%%%%%%%%%%%%%%%%%%%%%%%%%%%%%%%%%%%%%%%%%%%%%%%%%%%%%%%%%%%%%%%%%%%%%%%%
 \noindent Note that terms containing  ${\gamma^- p_f^+}$ never contribute to the trace because 
 there is always an adjacent factor of $\gamma^-$, and $\gamma^- \gamma^- = \gamma^+ \gamma^+ = 0$. 
 %
 % \checkmark %%%%%%%%%%%%%%%%%%%%%%%%%%%%%%%%%%%%%%%%%%%%%%%%%%%%%%%%%%%%%%%%%%%%%%%%%%%%%%%%%%%%%%%
One may evaluate the trace, using the relation $\gamma^+ \gamma^- = 2 - \gamma^- \gamma^+$ Eq.\eqref{nu11} simplifies to, 
 %
 % \checkmark %%%%%%%%%%%%%%%%%%%%%%%%%%%%%%%%%%%%%%%%%%%%%%%%%%%%%%%%%%%%%%%%%%%%%%%%%%%%%%%%%%%%%%%
 \begin{eqnarray}
 {{\cal N}_{11}^{c}} &=& \frac{2(2q^-)^3}{y}\left(1-y+y\eta\right)\left[P(y) l_\perp^2 + y^4 M^2\right].
 \end{eqnarray}
 %
 % \checkmark %%%%%%%%%%%%%%%%%%%%%%%%%%%%%%%%%%%%%%%%%%%%%%%%%%%%%%%%%%%%%%%%%%%%%%%%%%%%%%%%%%%%%%%
 Note that the mass independent portion contains the standard vacuum splitting function $P(y)$ while the mass
 dependent part has a separate dependence on the momentum fraction of the radiated gluon ($y$). 
 %
 % \checkmark %%%%%%%%%%%%%%%%%%%%%%%%%%%%%%%%%%%%%%%%%%%%%%%%%%%%%%%%%%%%%%%%%%%%%%%%%%%%%%%%%%%%%%%
In the soft emission kinematic limit where $y \ll 1$, one may neglect the mass term. However, in this work we will retain it 
 throughout. 

 The set of denominators can now be evaluated using contour integration. 
 For the central cut diagram of Fig.(\ref{Sample_Diagram}), this yields the phase factor,   
 % \checkmark %%%%%%%%%%%%%%%%%%%%%%%%%%%%%%%%%%%%%%%%%%%%%%%%%%%%%%%%%%%%%%%%%%%%%%%%%%%%%%%%%%%%%%%
 \begin{widetext} 
 \begin{eqnarray}
 {\bar I}_{11}^{c} &=& \exp\left[i\left(x_B+x_L+\frac{x_M}{1-y}\right)P^+\left({y'}^-_0-y_0^-\right)\right]
 \exp\left[i\left(\zeta x_D+\left(\zeta-1\right)\frac{x_M}{1-y}
 -\zeta\frac{\eta y^2}{1-y}x_L\right)P^+\left({y'}^-_1-y^-_1\right)\right] \nn \\
  &\times&\theta\left(y^-_1-y^-_0\right) \theta\left({y'}^-_1-{y'}^-_0\right)  \nn \\
  &\times&\left[1-\exp\left[-i\left(x_L+\frac{y}{1-y}x_M\right)P^+\left(y_1^- - y_0^- \right)\right]\right]
    \left[1-\exp\left[i\left(x_L+\frac{y}{1-y}x_M\right)P^+\left({y'}^-_1 - {y'}^-_0\right)\right]\right].  \nn 
 \end{eqnarray} 
 \end{widetext}
 This single diagram contains four contributions depending on the one propagator that remains off shell after 
 contour integration.

 Multiplying through we find four terms similar to the work of Ref.~\cite{Wang:2001ifa}.
 The first term where the same propagator is off-shell in both amplitude and complex conjugate corresponds 
 to the so called ``hard-soft'' process where the gluon radiation is induced by the initial hard scattering. 
 The heavy quark is knocked off-shell by the initial hard scattering and becomes on-shell after 
 radiating the on-shell gluon. 
 Afterwards, the on-shell quark or gluon will have a scattering with another soft medium gluon from the nucleus. 
 The second term is the case where the quark is on-shell immediately after the first hard scattering. 
 Gluon radiation is induced by subsequent scattering of the heavy quark off a in-medium gluon which 
 carries a specific finite momentum fraction. This is often referred to as ``hard-hard'' scattering. 
 The two cross terms where different propagators are off-shell in the amplitude and complex conjugate represent 
 interference between soft-hard and hard-hard scatterings.

 % \checkmark %%%%%%%%%%%%%%%%%%%%%%%%%%%%%%%%%%%%%%%%%%%%%%%%%%%%%%%%%%%%%%%%%%%%%%%%%%%%%%%%%%%%%%%
The equations derived above contain both longitudinal and transverse momentum exchanges with the medium. The portion due to transverse exchange may be isolated by imposing that $k^-\rightarrow 0~(\eta \rightarrow 0,~\zeta \rightarrow 1)$ limit, 
and then comparing with expressions from similar diagrams in Ref \cite{Zhang:2004qm}.
 %
 % \checkmark %%%%%%%%%%%%%%%%%%%%%%%%%%%%%%%%%%%%%%%%%%%%%%%%%%%%%%%%%%%%%%%%%%%%%%%%%%%%%%%%%%%%%%%
 One will immediately note that factors containing $x_M$ which 
 contribute to the Landau-Pomeranchuck-Migdal (LPM) \cite{Landau:1953um,Migdal:1956tc} effect 
 are not modified by factors of $k^-$, while factors containing $x_D$ are modified by the presence of $k^-$. 
 %
 % \checkmark %%%%%%%%%%%%%%%%%%%%%%%%%%%%%%%%%%%%%%%%%%%%%%%%%%%%%%%%%%%%%%%%%%%%%%%%%%%%%%%%%%%%%%%
 Factors of $x_D$ will eventually be absorbed in the definition of the transport coefficients including $\hat{q}$.
 Such factors introduce a non-trivial dependence of in-medium transport coefficients on the mass of the probe. 

In this section, we have demonstrated how a certain diagram for heavy quark production and energy loss via gluon radiation can be simplified. Similar rules will be applied to all other real diagrams which include a cut of the radiated gluon line. In the subsequent section, 
all real diagrams will be combined to obtain the real gluon emission spectrum from a heavy quark that undergoes one scattering and 
one emission after production.

 %\begin{widetext}

\section{gluon emission spectrum}

In the preceding section we evaluated the diagram in Fig.~\ref{Sample_Diagram}, in some detail, to highlight the 
approximations that will be made in the course of the full calculation. In this section, the result of the sum of 
all real diagrams (with an emitted gluon in the final state), will be presented. This will be followed by  a gradient 
expansion in the exchanged transverse momentum ($k_\perp \rightarrow 0$). 
While the leading term in the limit of $k_\perp \rightarrow 0$,  
will correspond to a gauge correction to the vacuum process of gluon radiation from a
heavy quark, the focus in this section will be on first correction 
in the $k_\perp \rightarrow 0$ limit, usually denoted as the next-to-leading twist contribution.

In total, there are 11 separate topologically distinct diagrams similar to that in Fig.~\ref{Sample_Diagram}. We denote these 
with two subscripts: $m,n = 1$-$3$, where, $m$ denotes the location of the scattering in the amplitude, and $n$ denotes the 
location in the complex conjugate, for the case of a central cut, where one gluon scattering is on either side of the cut line. In either case of $m$ or $n$,  
$1$ signifies that the scattering occurs on the quark line beyond emission, $2$ 
signifies scattering on the quark line between the hard production and the emission, and $3$ signifies scattering of the emitted gluon.  
Each one of these diagrams will also generate a left and right cut component, where the cut line will be moved to the left or right of the scatterings,  
with the topology of the diagram held fixed. There are also the two special configurations, where both scatterings occur between the hard production and 
the gluon emission in the amplitude or complex-conjugate. These are denoted as $\mathcal{C}_{0,1}$ and $\mathcal{C}_{1,0}$. 
The next-to-leading twist portion, of the sum over all three cuts, for each of these contributions is outlined in the appendix.

Adding all the contributions from all the diagrams, categorized in the Appendix, we obtain the entire contribution to the hadronic tensor. 
In what follows, we decompose the hadronic tensor as, %\note{
\bea
W^{\mu \nu} = g^{4} 2\pi ( - g_{\perp}^{\mu \nu} ) {\cal H}^{c,l,r}.
\eea
%}
The entire contribution from all real diagrams is contained in the factor  $\mathcal{H}^{c,l,r}$. This, includes the initial 
hard scattering, the final state scattering of the quark or gluon, and the emission vertex.  Virtual contributions, where the final state radiated 
gluon is not cut, will not be considered in this effort. Some part of the spin sum in the numerator has already been factorized out in the 
term $- g_{\perp}^{\mu \nu}$ above. In what follows, we will simplify 
${\cal H}^{c,l,r} $ by factoring different contributions within it and then applying approximations to them separately. In the interest of readability, the exact details of the calculation for each diagram separately is included in the appendix. 

This entire factor ${\cal H}^{c,l,r} $ is obtained as, 
 \begin{eqnarray}
  {\cal H}^{c,l,r}  
 &=& \sum_{m,n=1}^{3} {\cal C}_{m,n}^{c,l,r}  + \mathcal{C}_{0,1} + \mathcal{C}_{1,0}
 = \frac{2\pi \alpha_s}{N_c} \int d l_\perp^2 {H}^{c,l,r} \nn \\
  &\times& \exp\left[ i\left(x_B+x_L+\frac{x_M}{1-y}\right)P^+\left({y'}_0^--{y}_0^-\right)  \right. \nn \\
  &+& i\left(\zeta x_D+\left(\zeta-1\right)\frac{x_M}{1-y}
 -\zeta\frac{\eta y^2}{1-y}x_L\right) \nn \\
&\times& \left. P^+\left({y'}_1^--{y}_1^-\right) \right] \nn \\
&\times&  \langle A | {\bar \psi}(y_0') \gamma^+ A^{+}(y_1')  A^{+}(y_1) \psi(y_0) | A \rangle
 \end{eqnarray}
In the equation above, cut specific phase factors and the hard part for each cut are entirely contained within ${H}^{c,l,r}$.  
The overall phase factor represents the generic portion of the phase factor. 

 In order to
 calculate the next-to-leading power contribution to the semi-inclusive  hard partonic cross-section, one needs to expand the  cross-section
 in $k_\perp$ and in $k^-$.
In each case, we will extract the corresponding transport
 coefficients inside the gluon emission spectrum kernel for the semi-hard heavy quark
 \cite{Majumder:2009ge}.
 Factors of $k_\perp$ and $k^-$ are absorbed as derivatives within the definition of the transport coefficients 
 [e.g. $k_\perp A^+(\vec{y}_\perp) \exp(i \vec{k}_\perp \cdot \vec{y}_\perp) = -i \nabla_\perp A^+(\vec{y}_\perp) \exp(i \vec{k}_\perp \cdot \vec{y}_\perp) \simeq i F^{+\perp} (\vec{y}_\perp)  \exp(i \vec{k}_\perp \cdot \vec{y}_\perp)$].

 We will also factor the four point non-perturbative operator using the usual phenomenological factorization, which for the 
 case of transverse scattering may be expressed as, 
 \bea
&&  \langle A | {\bar \psi}(y_0') \gamma^+ F^{+}_\perp (y_1')  F^{+}_\perp (y_1) \psi(y_0) | A \rangle \simeq C^A_{p} \nn \\
 &\times& \langle p |   {\bar \psi}(y_0') \gamma^+ \psi(y_0) | p \rangle \frac{\rho}{2p^+} \langle p | F^{+}_\perp (y_1')  F^{+}_\perp (y_1) | p \rangle .
 \eea
 The first operator product on the right hand side of the equation above will yield the incoming quark distribution function within one nucleon. The second operator product will yield the transport coefficient due to the scattering of the final state, off a gluon within a nucleon in the nucleus. We have assumed the average condition that both nucleons have a momentum $p = P/A$. The factor $\rho$ represents the nucleon density within the nucleus, and $C_p^A$ represents an overall normalization constant that contains the nucleon density. The factor of $\rho/(2p^+)$ 
 is written separately as that will be absorbed within the definition of the transport coefficient.
 
 These transport coefficients, defined below, are non-perturbative objects, which are factorized from the hard part that describes the 
 propagation of the heavy quark. While the exact value of each of these coefficients depends on the non-perturbative dynamics of the 
 medium, the relative contributions of the different hard parts that appear as a multiplicative factor along with these coefficients will be 
 calculated below. 
 %
% Even though the transport coefficients are non perturbative objects their relative strength however calculable perturbatively.  
 Terms for the transverse diffusion $\hat{q}$, 
 the drag (and longitudinal diffusion) coefficient $\hat{e}$ ($\hat{e}_2$) can be obtained through derivatives of 
 the kernel with respect to the transverse and $(-)$-light-cone component of the exchange momentum, 
 \begin{eqnarray}
 &~&\left[\nabla_{k_\perp}^2, \nabla_{k^-}, \nabla_{k^-}^2 \right]\left.{H}^{c,l,r}\right|_{k_\perp, k^- = 0}  \nn \\
 &=& 4 C_A \left(\frac{1+(1-y)^2}{y}\right) \frac{l_\perp^4}{[l_\perp^2+y^2M^2]^4} {\tilde H}_{c,l,r}^{{\hat q},{\hat e},{\hat e}_2} .
 \end{eqnarray}

In the equation above, the factor ${\tilde H}_{c,l,r}^{{\hat q},{\hat e},{\hat e}_2} $ represents several terms, depending on the 
cut taken i.e., central $c$, left $l$, or right $r$, and the momentum component with respect to which the Taylor expansion is 
carried out, i.e., $\hat{q}$ for the second derivative in terms of $k_\perp$, $\hat{e}$ for the first derivative with respect to $k^-$  
and $\hat{e}_2$ for the second derivative with respect to $k^-$. One should note that for each case, once the derivatives have been 
taken, both factors of the momentum $k_\perp, k^-$ are set to zero. The complete expressions for $\tilde{H}_{c,l,r}^{{\hat q},{\hat e},{\hat e}_2}$ can be expressed as a sum of products of a phase factor and a non-phase factor coefficient, expressed as $c^{{\hat q},{\hat e},{\hat e}_2}_n$:
 \begin{widetext}
 \begin{eqnarray}
 {\tilde  H}_{c}^{{\hat q},{\hat e},{\hat e}_2} &=&  c_1^{{\hat q},{\hat e},{\hat e}_2}  \left[1-e^{-i\left(x_L+\frac{y}{1-y}x_M\right)P^+ \left({y}_1^--{y}_0^-\right)}\right]
    \left[1-e^{i\left(x_L+\frac{y}{1-y}x_M\right)P^+\left({y'}_1^--{y'}_0^-\right)}\right]
    + c_2^{{\hat q},{\hat e},{\hat e}_2}  \left\{ e^{-i\left(x_L+\frac{y}{1-y}x_M\right)P^+ \left({y}_1^--{y}_0^-\right)} \right. \nn \\        
   &\times&  \left. \left[1-e^{i\left(x_L+\frac{y}{1-y}x_M\right)P^+\left({y'}_1^--{y'}_0^-\right)}\right]  
     + \left[1-e^{-i\left(x_L+\frac{y}{1-y}x_M\right)P^+ \left({y}_1^--{y}_0^-\right)}\right]      
       e^{i\left(x_L+\frac{y}{1-y}x_M\right)P^+\left({y'}_1^--{y'}_0^-\right)}     \right\}                \nn \\
        &+& c_3^{{\hat q},{\hat e},{\hat e}_2}  \left[e^{-i\left(x_L+\frac{y}{1-y}x_M\right)P^+ \left({y}_1^--{y}_0^-\right)}\right] 
    \left[e^{i\left(x_L+\frac{y}{1-y}x_M\right)P^+\left({y'}_1^--{y'}_0^-\right)}\right], \nn \\ 
 {\tilde H}_{l}^{{\hat q},{\hat e},{\hat e}_2}  &=&  c_4^{{\hat q},{\hat e},{\hat e}_2}\left[e^{-i\left(x_L+\frac{y}{1-y}x_M\right)P^+\left({y'}_0^--{y'}_1^-\right)}
                     -e^{-i\left(x_L+\frac{y}{1-y}x_M\right)P^+\left({y'}_0^--{y}_1^-\right)}\right]   
         + c_5^{{\hat q},{\hat e},{\hat e}_2} \left[1-e^{-i\left(x_L+\frac{y}{1-y}x_M\right)P^+\left({y'}_0^--{y'}_1^-\right)}\right] ,\nn  \\ 
 {\tilde  H}_{r}^{{\hat q},{\hat e},{\hat e}_2}  &=& c_4^{{\hat q},{\hat e},{\hat e}_2}\left[e^{-i\left(x_L+\frac{y}{1-y}x_M\right)P^+\left({y}_0^--{y}_1^-\right)}
                     -e^{-i\left(x_L+\frac{y}{1-y}x_M\right)P^+\left({y}_0^--{y'}_1^-\right)}\right]   
         + c_5^{{\hat q},{\hat e},{\hat e}_2} \left[1-e^{-i\left(x_L+\frac{y}{1-y}x_M\right)P^+\left({y}_0^--{y}_1^-\right)}\right] . \nn
         \label{phase}
  \end{eqnarray}
  \end{widetext}
  In each case above, the coefficients $c_n^{{\hat q},{\hat e},{\hat e}_2}$ depend on the momentum component 
  being considered. The subscript $n$ merely denotes the order in which the coefficient occurs: $c_1$, $c_2$ and $c_3$ appear in the 
  expression for the central cut, where $c_4$ and $c_5$ appear in both the left and right cuts. We list them in the following for each different 
  case, starting from the case of transverse diffusion, i.e., $\hat{q}$. The coefficients are, 
 \begin{eqnarray}
 c_{1}^{\hat q}&=& 1 - (2-3\kappa y^2) \chi + (1-\kappa y^2) \chi^2  ~, \nn \\
 c_{2}^{\hat q}&=& -\frac{y}{2} +  \left[\left(1-\frac{1}{2}\kappa y^2 -\kappa y^3 \right) 
       +y^2\frac{C_F}{C_A}\left(2-\kappa y^2\right)\right]\chi  \nn \\
    &~&   -\left[\frac{1}{2}\left(y-\kappa y^2\right)-y^2\frac{C_F}{C_A}\left(\kappa y^2\right)\right]\chi^2 ~,\nn \\
 c_{3}^{\hat q}&=&y^2\frac{C_F}{C_A}\left[1-4(1-\kappa y^2)\chi+(1-2\kappa y^2)\chi^2\right] ~, \nn \\
 c_{4}^{\hat q} &=&  \left[1+\frac{y^2}{1+(1-y)^2}\chi\right]\left[\frac{C_F}{C_A}y^2+1-2y\right]\chi~,  \nn \\
 c_{5}^{\hat q} &=&  \left[1+ \frac{y^2}{{1+(1-y)^2}}\chi\right] \chi ~.
 \end{eqnarray}
 \noindent The momentum fractions $y$, $\kappa$ and $\chi$ are defined in Eq.~\eqref{momentum-fractions}. 
 For the longitudinal drag  coefficient $\hat{e}$, the $c$-factors are,  
 \begin{eqnarray}
 c_{1}^{\hat e}&=& \frac{y^2M^2}{l^-}\left[\frac{1}{2} + \frac{1}{2}(1+\kappa y^2)\chi+ \frac{1}{2} \chi^2 \right] ,    \nn \\
 c_{2}^{\hat e}&=& \frac{y^2M^2}{l^-}\left[-\frac{1}{4} - \frac{1}{4}(1+\kappa y^2)~y^2\chi - \frac{1}{4} \chi^2 \right]  ,          \nn \\
 c_{3}^{\hat e}&=& 0~,
 c_{4}^{\hat e} = 0~, 
 c_{5}^{\hat e} = 0. \nn 
 \end{eqnarray}
 
 \noindent For longitudinal diffusion coefficient $\hat{e}_2$, the $c$-factors are, 
 \begin{eqnarray}
 c_{1}^{{\hat e}_2}&=& \frac{y^2M^2}{(l^-)^2}\left[-\frac{1}{2} + \left(\frac{7}{2}-\frac{1}{2}\kappa y^2\right)\chi 
 + \frac{7}{2}~y^2\chi^2   \right] , \nn \\
 c_{2}^{{\hat e}_2}&=&  
      \frac{y^2M^2}{(l^-)^2} \left[\frac{1}{4} - \left(\frac{3}{4}-\frac{1}{4}\kappa y^2\right)\chi- 
      \frac{3}{4} \kappa y^2 ~ \chi^2\right] ,            \nn \\
 c_{3}^{{\hat e}_2}&=& 0~,~~
 c_{4}^{{\hat e}_2} = 0~, ~~
 c_{5}^{{\hat e}_2} = 0~. \nn 
 \end{eqnarray}

All the terms presented above can be combined to obtain the real single gluon emission spectrum. 
 In the second line of the equation below [Eq. \eqref{master}], we have retained terms only up to
 ${\cal O}(\sqrt{\lambda})$, the approximation that has been justified in this study. 
All terms which scale as ${\cal O}(\lambda)$ or greater, have been neglected. We express the gluon spectrum per 
unit light-cone length as, 
 \begin{widetext}
 \begin{eqnarray}
 \frac{dN_g}{dy dl^2_\perp d\tau} &=&2~\frac{\alpha}{\pi}~P(y)~\frac{1}{l^4_\perp} \left(\frac{1}{1+\chi}\right)^{4}
                 ~\sin^2\left(\frac{l^2_\perp}{4l^-(1-y)}(1+\chi)~\tau\right)\left[\left\{c_1^{\hat q}+c_2^{\hat q}\right\}~{\hat q}
           +4\left\{c_1^{\hat e}+c_2^{\hat e}\right\}~{\hat e}
           +2\left\{c_1^{{\hat e}_2}+c_2^{{\hat e}_2}\right\}~{\hat e_2}\right] \nn \\
             &=&2~\frac{\alpha}{\pi}~P(y)~\frac{1}{l^4_\perp} \left(\frac{1}{1+\chi}\right)^{4}
                 ~\sin^2\left(\frac{l^2_\perp}{4l^-(1-y)}(1+\chi)~\tau\right)  \nn \\
          & \times& \left[\left\{\left(1-\frac{y}{2}\right)-\chi
               +\left(1-\frac{y}{2}\right)\chi^2\right\}~{\hat q}
           +\frac{l_\perp^2}{l^-}\chi\left(1+\chi\right)^2~{\hat e}
           +\frac{l_\perp^2}{(l^-)^2}\chi\left(\frac{1}{2}-\frac{11}{4}\chi\right){\hat e_2}\right] .   \label{master} 
 \end{eqnarray}
 \end{widetext}
 In the equation above, we have defined a mean light-cone location of the first scattering (between the amplitude and complex conjugate) as
 %\note{
 \bea
 \tau = \frac{y_1^- + {y'}_1^-}{2}.
\eea
We also define the off-set between the light cone locations in the amplitude and complex-conjugate as, 
\bea
y^- = y_1^- - {y'}_1^-.
\eea
%}
This variable enters the definitions of all transport coefficients that will be discussed in this paper. 
There are three transport coefficients, which contain 
the non-perturbative expection of the gluon field strength operators: the transverse diffusion coefficient, $\hat{q}$ which represents the 
transverse momentum squared per unit light-cone length, exchanged between the hard quark and the medium, the longitudinal drag per unit light-cone length, $\hat{e}$ caused due to the exchange of light-cone components of momentum, and $\hat{e}_2$ the diffusion in light-cone 
momentum, per unit 
light-cone length : 
 \begin{eqnarray}
 {\hat q} &=& \frac{4\pi^2 C_R\alpha_s}{N_C^2-1}\int \frac{dy^-}{\pi} \frac{\rho}{2p^+}  \nn \\
        &\times&   \langle A | F_\perp^{~+}(y^-) F^{\perp+}(0)| A\rangle~e^{-i{\bar \Delta}P^+y^-} , \nn \\
 {\hat e} &=& \frac{4\pi^2 C_R\alpha_s}{N_C^2-1}\int \frac{dy^-}{\pi} \frac{\rho}{2p^+}  \nn \\
        &\times&  \langle A | i \partial^{-} A^+(y^-) A^+(0)| A\rangle~e^{-i{\bar \Delta}P^+y^-} , \nn \\
 {\hat e_2} &=& \frac{4\pi^2 C_R\alpha_s}{N_C^2-1}\int \frac{dy^-}{\pi} \frac{\rho}{2p^+}  \nn \\
        &\times&          \langle A | F^{-+}(y^-) F^{-+}(0)| A\rangle~e^{-i{\bar \Delta}P^+y^-} .
 \end{eqnarray}
 \noindent In the equations above, we observe the appearance of another momentum fraction:
 \begin{eqnarray}
 \bar \Delta &=& \zeta x_D+\left(\zeta-1\right)\frac{x_M}{1-y}
 -\zeta\frac{\eta y^2}{1-y}x_L~.
 \end{eqnarray}
 The presence of such a momentum fraction, indicates that the range of momentum fractions in the definition of $\hat{q}$
$\hat{e}$ and $\hat{e}_2$ for heavy quark scattering is different from that for light flavor energy loss. This indicates that the 
actual value of $\hat{q}$ (or even $\hat{e}$ or $\hat{e}_2$) for heavy quarks may be different from that for light quarks.
Thus, careful analysis of heavy-quark energy loss may lead to an understanding of the $x$-dependence of the in-medium 
gluon distribution function that sources transport coefficients, and may, in the end, lead to an understanding of the degrees of freedom within 
dense media, where heavy quark energy loss is carried out.

 \section{Conclusion and Outlook}
 In this work, the gluon bremsstrahlung from a  ``semi-hard" heavy quark in a dense nuclear medium has been studied in greater 
 detail than in several earlier efforts. 
In this work, we have considered a hard virtual photon scattering off a hard heavy quark (within a proton),  that converts it to a slow moving heavy quark that 
moves through the remainder of the nucleus before escaping and fragmenting into a jet containing a heavy meson. 

 In this work both transverse broadening as well as the longitudinal drag and longitudinal diffusion, have been
 studied on an equal footing.
 We have categorically focussed our study on ``semi-hard'' quarks where the mass and momentum scale as
 $M, p \sim \sqrt{\lambda} Q$, as these are the quarks for which mass modifications is most prominent. 
 We have used power counting arguments loosely based on Soft Collinear Effective Theory (SCET) at various stage to isolate the
 leading contributions.
 It was shown in our earlier studies that both longitudinal and transverse momentum transfers have a
 comparable effect on the off-shellness of the heavy-quark \cite{Abir:2014sxa}.
 This earlier work implied that longitudinal transfers, not only lead to the drag and diffusion, similar to
 light flavors, but will also noticeably affect the radiative loss and left strong indications
 that for heavy quarks, the drag induced radiation may be as significant as transverse momentum diffusion ($\hat{q}$) induced radiation.
 
 In this paper we have explicitly demonstrated that the gluon bremsstrahlung spectrum from a semi-hard heavy quark 
 is indeed strongly modified by drag induced radiation.
% for heavy quarks.
 %
 We have shown that due to the presence of the $(-)$-light-cone momentum exchange from the medium 
% factors belongs to jet quenching parameter modifies non trivially due to non-negligible
 ($k^-$), in our calculations, the definition of all the transport coefficients for heavy quark is different from that for light quark. 
 %
% This clearly shows how those non-perturbative jet transport quantities may depend though weakly on
 Thus transport coefficients may indeed depend on properties of the probe {\it i.e.} mass or on the $l_\perp^2$. Whether this is phenomenologically significant cannot be ascertained at this point, and is left for a future investigation.
 This explicit dependence on the $(-)$-light cone momentum was absent in the limit of $k^-\rightarrow 0$, assumed in several prior calculations.

 The implications of the present study on the  phenomenology of HIC is under way.  
 It this work we have shown that the gluon bremsstrahlung spectrum of heavy quark (unlike light quark)
 is parametrically sensitive to $\hat e$ which quantifies the amount of drag the moving quark experiences.
 This result can be used to estimate the value of this sub-leading non-perturbative jet transport
 parameter ($\hat e$) from heavy flavor data of HIC experiments. These extra additive contributions may lead to an eventual solution of the heavy quark puzzle. 
 We leave these for a future effort. 
% which have been left for future. 

 \section{Acknowledgments}
 The authors would like to thank G.-Y. Qin for many helpful discussions.
 This work was supported in part by the National Science Foundation under grant number PHY-1207918. 
 This work is also supported in part by the Director, Office of Energy Research, Office of High Energy and 
 Nuclear Physics, Division of Nuclear Physics, of the U.S. Department of Energy, through the JET topical collaboration.

  \newpage

 \begin{widetext}
 
 \section{Appendix}

 % \checkmark %%%%%%%%%%%%%%%%%%%%%%%%%%%%%%%%%%%%%%%%%%%%%%%%%%%%%%%%%%%%%%%%%%%%%%%%%%%%%%%%%%%%%%%
  \begin{figure}[ht]
   \begin{center}
 \includegraphics[width=8cm,height=4cm]{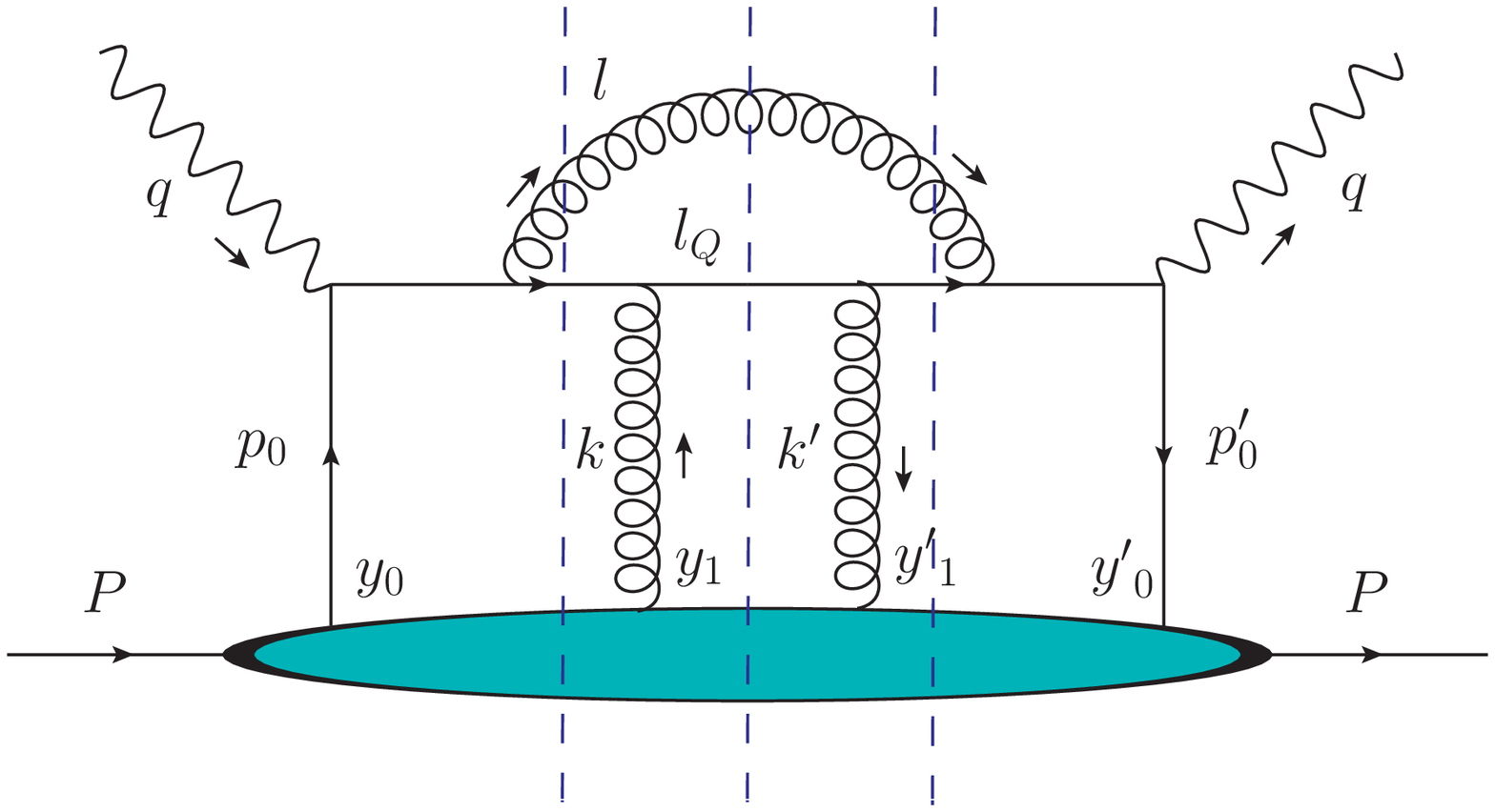}
 \caption{}
      \label{mult-scat}
  \end{center}
 \end{figure}
 %
 % \checkmark %%%%%%%%%%%%%%%%%%%%%%%%%%%%%%%%%%%%%%%%%%%%%%%%%%%%%%%%%%%%%%%%%%%%%%%%%%%%%%%%%%%%%%%
 
 There are three different cuts (central, left and right) in Fig.[\ref{mult-scat}] and their contributions are, 
 
 \begin{eqnarray}
 {\cal C}_{11}^{c,l,r} &=& \frac{\alpha_s^2}{N_c}\left[C_F\right]
                   \int dl_\perp^2 \left(\frac{1+(1-y)^2}{y}\right)
                   \frac{[l_\perp^2+ \kappa y^4M^2]}{(l_\perp^2+y^2M^2)^2} 
                   ~ {\bar I}_{11}^{c,l,r}
 \end{eqnarray}
 %
 % \checkmark %%%%%%%%%%%%%%%%%%%%%%%%%%%%%%%%%%%%%%%%%%%%%%%%%%%%%%%%%%%%%%%%%%%%%%%%%%%%%%%%%%%%%%%
 \begin{eqnarray}
 {\bar I}_{11}^{c} &=& \exp\left[i\left(x_B+x_L+\frac{x_M}{1-y}\right)P^+\left({y'}_0^--{y}_0^-\right)\right]
 \exp\left[i\left(\zeta x_D+\left(\zeta-1\right)\frac{x_M}{1-y}
 -\zeta\frac{\eta y^2}{1-y}x_L\right)P^+\left({y'}_1^--{y}_1^-\right)\right] \theta\left({y}_0^--{y}_1^-\right)  \nn \\
 && \times ~ \theta\left({y'}_0^--{y'}_1^-\right) \left[1-\exp\left[-i\left(x_L+\frac{y}{1-y}x_M\right)P^+
 \left({y}_1^--{y}_0^-\right)\right]\right]
    \left[1-\exp\left[i\left(x_L+\frac{y}{1-y}x_M\right)P^+\left({y'}_1^--{y'}_0^-\right)\right]\right]   \nn \\
 \end{eqnarray}
 \begin{eqnarray}
 {\bar I}_{11}^{l} &=& \exp\left[i\left(x_B+x_L+\frac{x_M}{1-y}\right)P^+\left({y'}_0^--{y}_0^-\right)\right]
 \exp\left[i\left(\zeta x_D+\left(\zeta-1\right)\frac{x_M}{1-y}
 -\zeta\frac{\eta y^2}{1-y}x_L\right)P^+\left({y'}_1^--{y}_1^-\right)\right] \nn \\
 && \times ~ \theta\left({y'}_1^--{y}_1^-\right) \theta\left({y'}_0^--{y'}_1^-\right) \left(-1\right)
    \left[1-\exp\left[i\left(x_L+\frac{y}{1-y}x_M\right)P^+\left({y'}_1^--{y'}_0^-\right)\right]\right]   
  \end{eqnarray}
 \begin{eqnarray}
 {\bar I}_{11}^{c} &=& \exp\left[i\left(x_B+x_L+\frac{x_M}{1-y}\right)P^+\left({y'}_0^--{y}_0^-\right)\right]
 \exp\left[i\left(\zeta x_D+\left(\zeta-1\right)\frac{x_M}{1-y}
 -\zeta\frac{\eta y^2}{1-y}x_L\right)P^+\left({y'}_1^--{y}_1^-\right)\right] \theta\left(y_0^--y_1^-\right)  \nn \\
 && \times ~ \theta\left({y}_1^--{y'}_1^-\right) \left[1-\exp\left[-i\left(x_L+\frac{y}{1-y}x_M\right)P^+
 \left({y}_1^--{y}_0^-\right)\right]\right]
    \left(-1\right) 
 \end{eqnarray}
 % \checkmark %%%%%%%%%%%%%%%%%%%%%%%%%%%%%%%%%%%%%%%%%%%%%%%%%%%%%%%%%%%%%%%%%%%%%%%%%%%%%%%%%%%%%%%
  \begin{figure}[ht]
   \begin{center}
 \includegraphics[width=8cm,height=4cm]{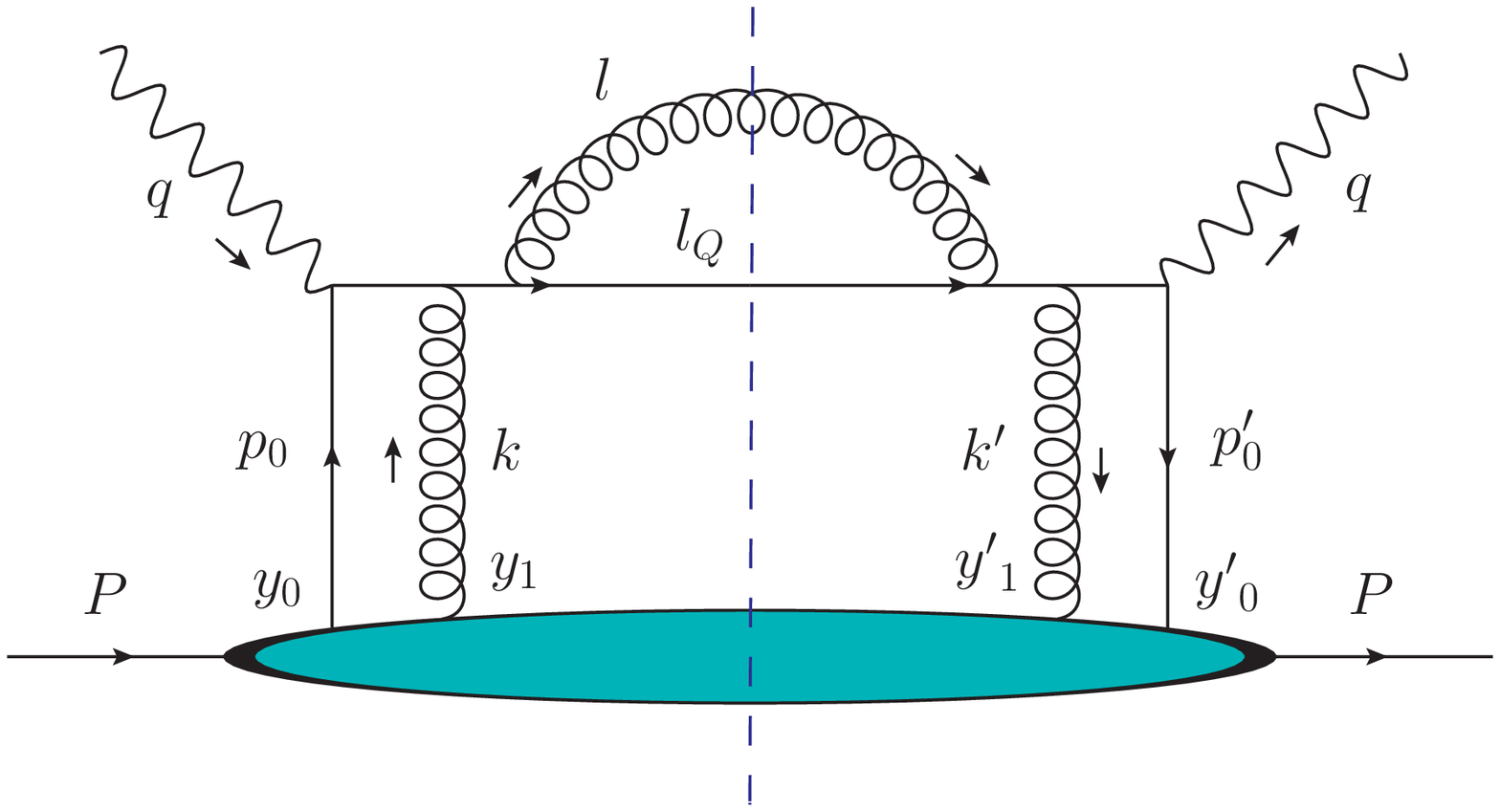}
 \caption{}
      \label{base_22}
  \end{center}
 \end{figure}
 %
 %
 % \checkmark %%%%%%%%%%%%%%%%%%%%%%%%%%%%%%%%%%%%%%%%%%%%%%%%%%%%%%%%%%%%%%%%%%%%%%%%%%%%%%%%%%%%%%%
 In Fig.[\ref{base_22}] there are only central cut, with the contribution, 
 
 \begin{eqnarray}
 {\cal C}_{22}^{c} &=& \frac{\alpha_s^2}{N_c}\left[C_F\right]
                   \int dl_\perp^2 \left(\frac{1+\eta y +(1-y+\eta y)^2}{y}\right)
                   \frac{[\left[(1+\eta y)l_\perp-yk_\perp\right]^2+ \kappa y^4M^2]}
                  {\left[\left(l_\perp-yk_\perp\right)^2+y^2M^2+2y\eta \left(l_\perp^2-l_\perp k_\perp\right)
                  +y^2\eta^2 l_\perp^2\right]^2}
                   ~ {\bar I}_{11}^{c,l,r}
 \end{eqnarray}
 %
 % \checkmark %%%%%%%%%%%%%%%%%%%%%%%%%%%%%%%%%%%%%%%%%%%%%%%%%%%%%%%%%%%%%%%%%%%%%%%%%%%%%%%%%%%%%%%
 \begin{eqnarray}
 {\bar I}_{22}^{c} &=& \exp\left[i\left(x_B+x_L+\frac{x_M}{1-y}\right)P^+\left({y'}_0^--{y}_0^-\right)\right]
 \exp\left[i\left(\zeta x_D+\left(\zeta-1\right)\frac{x_M}{1-y}
 -\zeta\frac{\eta y^2}{1-y}x_L\right)P^+\left({y'}_1^--{y}_1^-\right)\right] \theta\left({y}_0^--{y}_1^-\right)  \nn \\
 && \times ~ \theta\left({y'}_0^--{y'}_1^-\right) \left[-\exp\left[-i\left(x_L+\frac{y}{1-y}x_M\right)P^+
 \left({y}_1^--{y}_0^-\right)\right]\right]
    \left[-\exp\left[i\left(x_L+\frac{y}{1-y}x_M\right)P^+\left({y'}_1^--{y'}_0^-\right)\right]\right]
 \end{eqnarray}
 %
 % \checkmark %%%%%%%%%%%%%%%%%%%%%%%%%%%%%%%%%%%%%%%%%%%%%%%%%%%%%%%%%%%%%%%%%%%%%%%%%%%%%%%%%%%%%%%
  \begin{figure}[ht]
   \begin{center}
 \includegraphics[width=8cm,height=4cm]{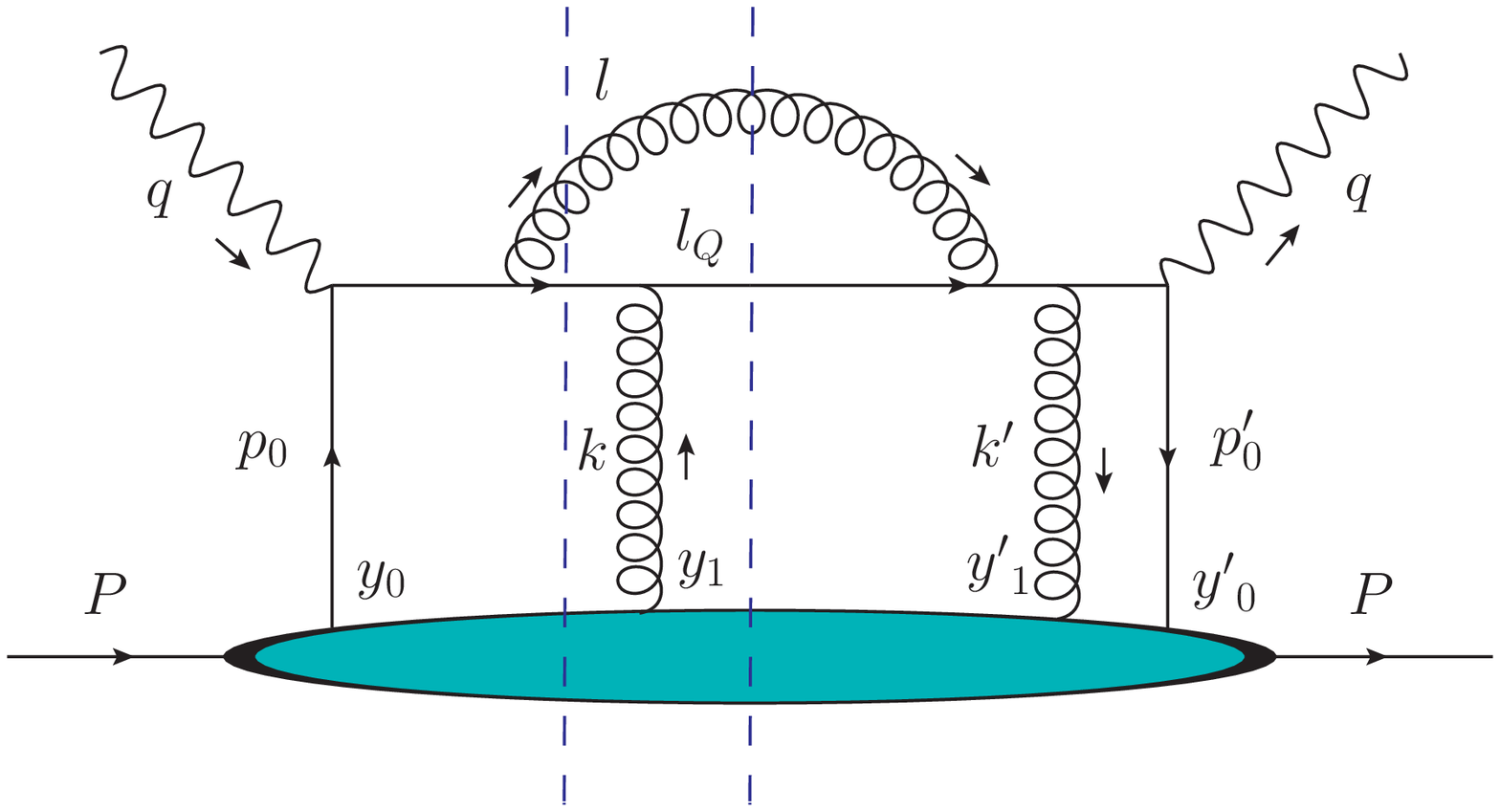}
 \caption{}
      \label{base_12}
  \end{center}
 \end{figure}
  
  In Fig.[\ref{base_12}] there are two different cuts for induced gluon radition, central and left. 
 
 \begin{eqnarray}
 {\cal C}_{12}^{c,l} &=& \frac{\alpha_s^2}{N_c}\left[-\left(C_F-\frac{C_A}{2}\right)\right]
                  \int dl_\perp^2 \left(\frac{1+(1-y)^2+\eta y (2-y)}{y}\right) \nn \\
                 && \times~ \frac{[l_\perp\left[(1+\eta y)l_\perp-yk_\perp\right]+ \kappa y^4M^2]}
                  {[l_\perp^2+y^2M^2]
                  \left[\left(l_\perp-yk_\perp\right)^2+y^2M^2+2y\eta \left(l_\perp^2-l_\perp k_\perp\right)
                  +y^2\eta^2 l_\perp^2\right]} {\bar I}_{12}^{c,l} 
 \end{eqnarray}
  % \checkmark %%%%%%%%%%%%%%%%%%%%%%%%%%%%%%%%%%%%%%%%%%%%%%%%%%%%%%%%%%%%%%%%%%%%%%%%%%%%%%%%%%%%%%%
 \begin{eqnarray}
 {\bar I}_{12}^{c} &=& \exp\left[i\left(x_B+x_L+\frac{x_M}{1-y}\right)P^+\left({y'}_0^--{y}_0^-\right)\right]
 \exp\left[i\left(\zeta x_D+\left(\zeta-1\right)\frac{x_M}{1-y}
 -\zeta\frac{\eta y^2}{1-y}x_L\right)P^+\left({y'}_1^--{y}_1^-\right)\right] \theta\left({y}_0^--{y}_1^-\right)  \nn \\
 && \times ~ \theta\left({y'}_0^--{y'}_1^-\right) \left[1-\exp\left[-i\left(x_L+\frac{y}{1-y}x_M\right)P^+
 \left({y}_1^--{y}_0^-\right)\right]\right]
    \left[-\exp\left[i\left(x_L+\frac{y}{1-y}x_M\right)P^+\left({y'}_1^--{y'}_0^-\right)\right]\right]   \nn \\
  \end{eqnarray}
 \begin{eqnarray}
 {\bar I}_{12}^{l} &=& \exp\left[i\left(x_B+x_L+\frac{x_M}{1-y}\right)P^+\left({y'}_0^--{y}_0^-\right)\right]
 \exp\left[i\left(\zeta x_D+\left(\zeta-1\right)\frac{x_M}{1-y}
 -\zeta\frac{\eta y^2}{1-y}x_L\right)P^+\left({y'}_1^--{y}_1^-\right)\right] \nn \\
 && \times ~ \theta\left({y'}_1^--{y}_1^-\right) \theta\left({y'}_0^--{y'}_1^-\right) 
    \left[\exp\left[i\left(x_L+\frac{y}{1-y}x_M\right)P^+
 \left({y}_1^--{y'}_0^-\right)+i\left(x_K-x_D\right)P^+\left({y'}_1^--y_1^-\right)\right] \right. \nn \\
 && \left. -\exp\left[i\left(x_L+\frac{y}{1-y}x_M\right)P^+\left({y'}_1^--{y'}_0^-\right)\right]\right]   
 \end{eqnarray}
 %
 %
  % \checkmark %%%%%%%%%%%%%%%%%%%%%%%%%%%%%%%%%%%%%%%%%%%%%%%%%%%%%%%%%%%%%%%%%%%%%%%%%%%%%%%%%%%%%%%
  \begin{figure}[ht]
   \begin{center}
 \includegraphics[width=8cm,height=4cm]{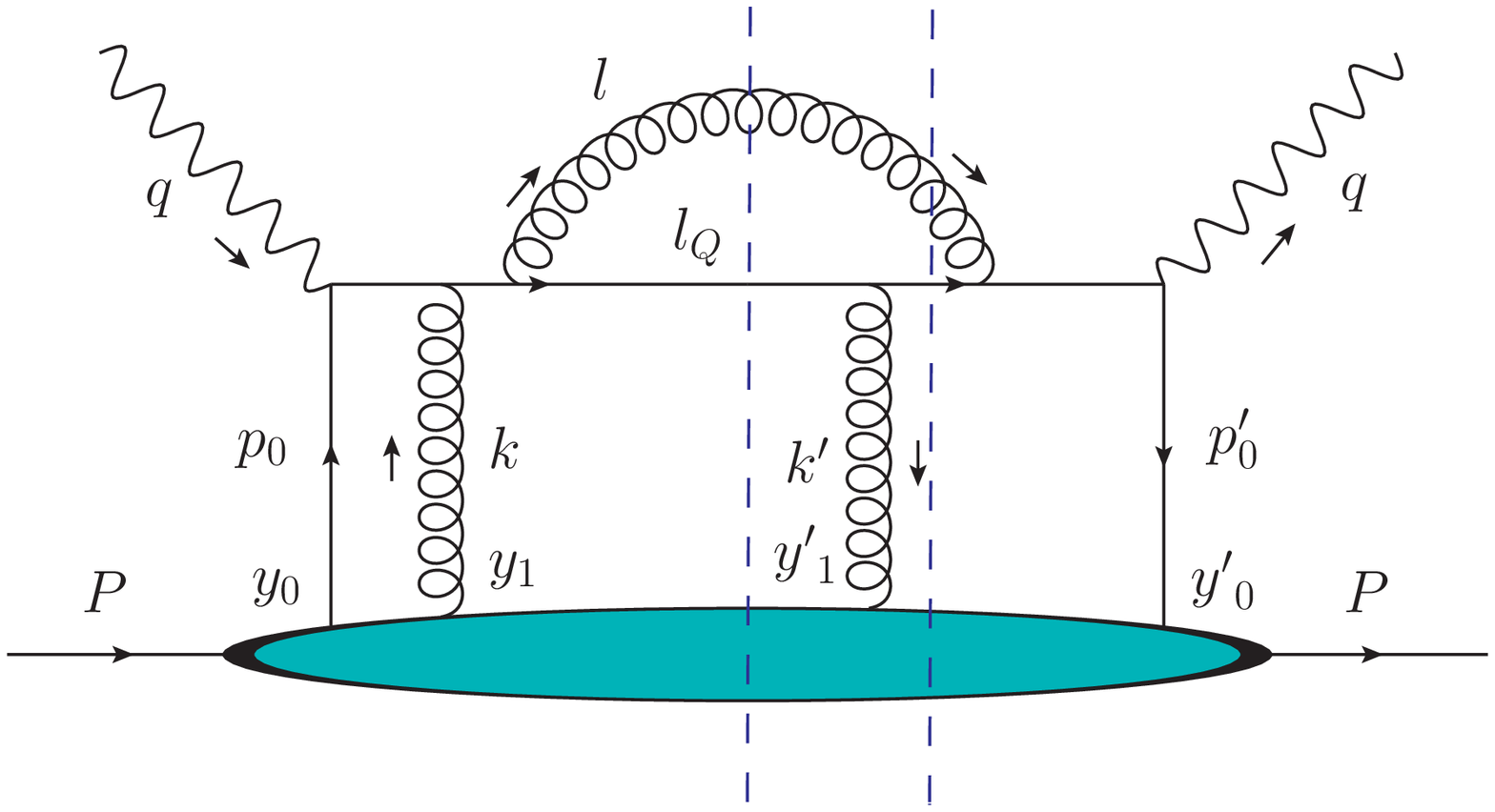}
 \caption{}
      \label{base_21}
  \end{center}
 \end{figure}

 In Fig.[\ref{base_21}] there are two different cuts for induced gluon radition, central and right.

 \begin{eqnarray}
 {\cal C}_{21}^{c,r} &=& \frac{\alpha_s^2}{N_c}\left[-\left(C_F-\frac{C_A}{2}\right)\right]
                  \int dl_\perp^2 \left(\frac{1+(1-y)^2+\eta y (2-y)}{y}\right) \nn \\
                 && \times~ 
                  \frac{[\left[(1+\eta y)l_\perp-yk_\perp\right]l_\perp+ \kappa y^4M^2]}
                  {\left[\left(l_\perp-yk_\perp\right)^2+y^2M^2+2y\eta \left(l_\perp^2-l_\perp k_\perp\right)
                  +y^2\eta^2 l_\perp^2\right][l_\perp^2+y^2M^2]} \nn    {\bar I}_{21}^{c,r} 
 \end{eqnarray}
 % \checkmark %%%%%%%%%%%%%%%%%%%%%%%%%%%%%%%%%%%%%%%%%%%%%%%%%%%%%%%%%%%%%%%%%%%%%%%%%%%%%%%%%%%%%%%
 \begin{eqnarray}
 {\bar I}_{21}^{c} &=& \exp\left[i\left(x_B+x_L+\frac{x_M}{1-y}\right)P^+\left({y'}_0^--{y}_0^-\right)\right]
 \exp\left[i\left(\zeta x_D+\left(\zeta-1\right)\frac{x_M}{1-y}
 -\zeta\frac{\eta y^2}{1-y}x_L\right)P^+\left({y'}_1^--{y}_1^-\right)\right] \theta\left({y}_0^--{y}_1^-\right)  \nn \\
 && \times ~ \theta\left({y'}_0^--{y'}_1^-\right) \left[-\exp\left[-i\left(x_L+\frac{y}{1-y}x_M\right)P^+
 \left({y}_1^--{y}_0^-\right)\right]\right]
    \left[1-\exp\left[i\left(x_L+\frac{y}{1-y}x_M\right)P^+\left({y'}_1^--{y'}_0^-\right)\right]\right]   \nn \\
 \end{eqnarray}
 \begin{eqnarray}
 {\bar I}_{21}^{r} &=& \exp\left[i\left(x_B+x_L+\frac{x_M}{1-y}\right)P^+\left({y'}_0^--{y}_0^-\right)\right]
 \exp\left[i\left(\zeta x_D+\left(\zeta-1\right)\frac{x_M}{1-y}
 -\zeta\frac{\eta y^2}{1-y}x_L\right)P^+\left({y'}_1^--{y}_1^-\right)\right] \nn \\
 && \times ~ \theta\left({y}_1^--{y'}_1^-\right) \theta\left({y}_0^--{y}_1^-\right) 
    \left[\exp\left[-i\left(x_L+\frac{y}{1-y}x_M\right)P^+
 \left({y'}_1^--{y}_0^-\right)-i\left(x_K-x_D\right)P^+\left({y}_1^--{y'}_1^-\right)\right] \right. \nn \\
 && \left. -\exp\left[-i\left(x_L+\frac{y}{1-y}x_M\right)P^+\left({y}_1^--{y}_0^-\right)\right]\right]   
 \end{eqnarray}
 %
 % \checkmark %%%%%%%%%%%%%%%%%%%%%%%%%%%%%%%%%%%%%%%%%%%%%%%%%%%%%%%%%%%%%%%%%%%%%%%%%%%%%%%%%%%%%%%
  \begin{figure}[ht]
   \begin{center}
 \includegraphics[width=8cm,height=4cm]{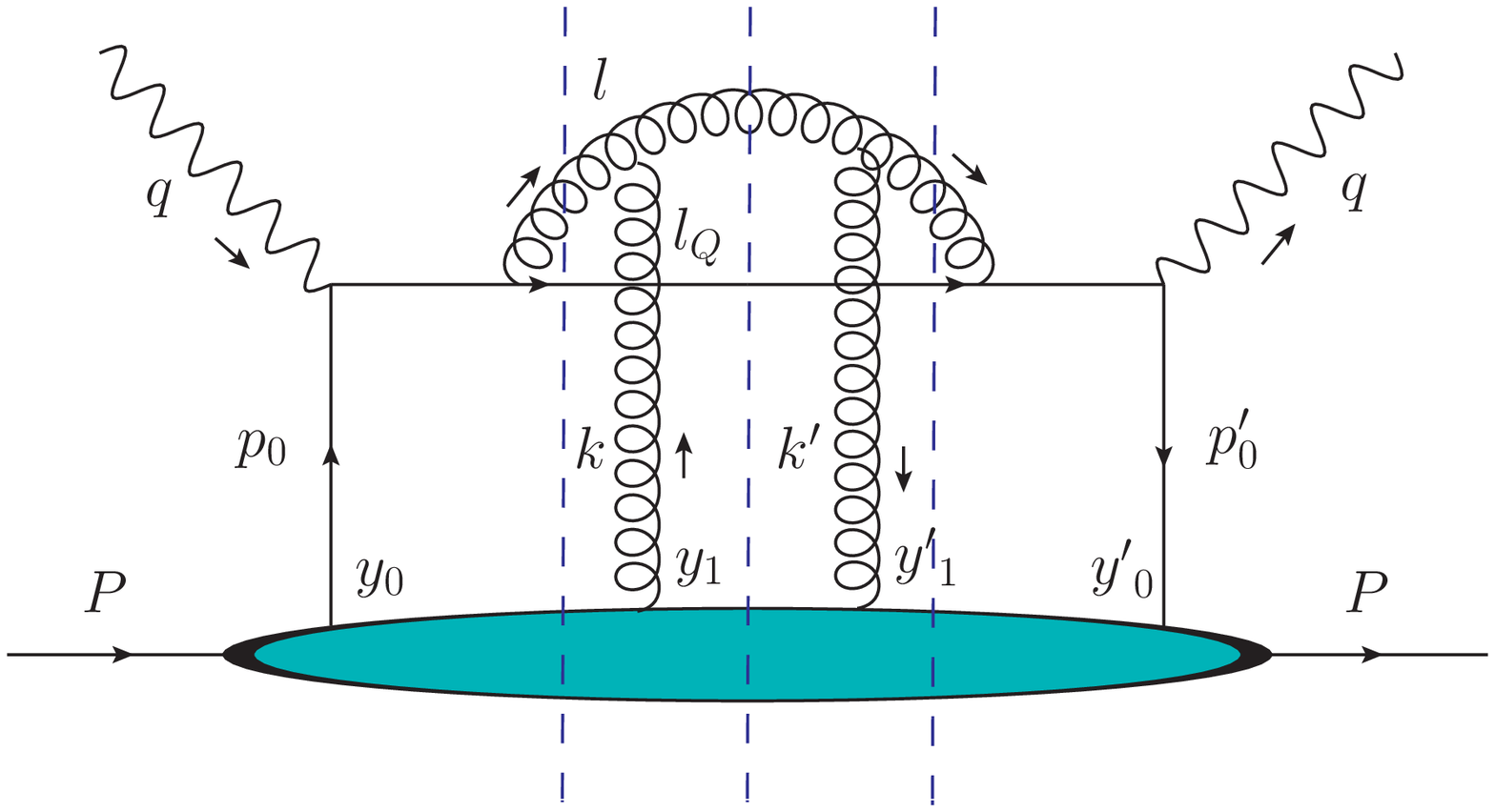}
 \caption{}
      \label{base_33}
  \end{center}
 \end{figure}

 In Fig.[\ref{base_33}] there are again all three possible cuts for induced gluon radition,

 \begin{eqnarray}
 {\cal C}_{33}^{c} &=& \frac{\alpha_s^2}{N_c}\left[C_A\right]
                  \int dl_\perp^2 P(y)
                  \frac{[\left(l_\perp-k_\perp\right)^2+ \kappa y^4M^2]}
                  {[\left(l_\perp-k_\perp\right)^2+\left(1-\eta\right)^2y^2M^2]^2} \nn {\bar I}_{33}^{c}
 \end{eqnarray}
 %
 % \checkmark %%%%%%%%%%%%%%%%%%%%%%%%%%%%%%%%%%%%%%%%%%%%%%%%%%%%%%%%%%%%%%%%%%%%%%%%%%%%%%%%%%%%%%%
 \begin{eqnarray}
 {\bar I}_{33}^{c} &=& \exp\left[i\left(x_B+x_L+\frac{x_M}{1-y}\right)P^+\left({y'}_0^--{y}_0^-\right)\right]
 \exp\left[i\left(\zeta x_D+\left(\zeta-1\right)\frac{x_M}{1-y}
 -\zeta\frac{\eta y^2}{1-y}x_L\right)P^+\left({y'}_1^--{y}_1^-\right)\right] \theta\left({y}_0^--{y}_1^-\right)  \nn \\
 && \times ~ \theta\left({y'}_0^--{y'}_1^-\right) \left[\exp\left\{i\left(\frac{\zeta}{y(1-\eta)}x_D
 +\frac{\eta(1-y)}{\left(1-\eta\right)}x_L
 +\left(\zeta-1\right)\frac{x_M}{1-y}
 -\zeta\frac{\eta y^2}{1-y}x_L\right)P^+\left(y_1^--y_0^-\right)
 \right\} \right. \nn \\
 && \left. -\exp\left[-i\left(x_L+\frac{y}{1-y}x_M\right)P^+
 \left({y}_1^--{y}_0^-\right)\right]\right] \left[\exp\left\{-i\left(\frac{\zeta}{y(1-\eta)}x_D
 +\frac{\eta(1-y)}{\left(1-\eta\right)}x_L
 +\left(\zeta-1\right)\frac{x_M}{1-y}
 -\zeta\frac{\eta y^2}{1-y}x_L\right) \right. \right. \nn \\
 && \left. \left. \times ~ P^+\left({y'}_1^--{y'}_0^-\right)
 \right\}-\exp\left[i\left(x_L+\frac{y}{1-y}x_M\right)P^+\left({y'}_1^--{y'}_0^-\right)\right]\right]    
 \end{eqnarray}
 
  \begin{eqnarray}
 {\cal C}_{33}^{l,r} &=& \frac{\alpha_s^2}{N_c}\left[C_A\right]
                  \int dl_\perp^2 P(y)
                  \frac{l_\perp^2+\kappa y^4 M^2}
                  {[l_\perp^2+y^2M^2]^2} \nn {\bar I}_{33}^{l,r}
 \end{eqnarray}
 %
 % \checkmark %%%%%%%%%%%%%%%%%%%%%%%%%%%%%%%%%%%%%%%%%%%%%%%%%%%%%%%%%%%%%%%%%%%%%%%%%%%%%%%%%%%%%%%
 \begin{eqnarray}
 {\bar I}_{33}^{l} &=& \exp\left[i\left(x_B+x_L+\frac{x_M}{1-y}\right)P^+\left({y'}_0^--{y}_0^-\right)\right]
 \exp\left[i\left(\zeta x_D+\left(\zeta-1\right)\frac{x_M}{1-y}
 -\zeta\frac{\eta y^2}{1-y}x_L\right)P^+\left({y'}_1^--{y}_1^-\right)\right] \theta\left({y'}_1^--{y}_1^-\right)  \nn \\
 && \times ~ \theta\left({y'}_0^--{y'}_1^-\right) \left[-\exp\left\{i\left(\frac{-\zeta(1-2y(1-\eta))}{y(1-\eta)}x_D
 -\frac{\eta(1-y)}{\left(1-\eta\right)}x_L
 +\left(\zeta-1\right)\frac{x_M}{1-y}
 -\zeta\frac{\eta y^2}{1-y}x_L\right)P^+\left(y_1^--{y'}_1^-\right)
 \right\} \right]  \nn \\
 && \left[1-\exp\left[i\left(x_L+\frac{y}{1-y}x_M\right)P^+\left({y'}_1^--{y'}_0^-\right)\right]\right]   
  \end{eqnarray}
 \begin{eqnarray}
 {\bar I}_{33}^{r} &=& \exp\left[i\left(x_B+x_L+\frac{x_M}{1-y}\right)P^+\left({y'}_0^--{y}_0^-\right)\right]
 \exp\left[i\left(\zeta x_D+\left(\zeta-1\right)\frac{x_M}{1-y}
 -\zeta\frac{\eta y^2}{1-y}x_L\right)P^+\left({y'}_1^--{y}_1^-\right)\right] \theta\left({y}_0^--{y}_1^-\right)  \nn \\
 && \times ~ \theta\left({y}_1^--{y'}_1^-\right) \left[1 -\exp\left[-i\left(x_L+\frac{y}{1-y}x_M\right)P^+
 \left({y}_1^--{y}_0^-\right)\right]\right]   \nn \\
 && \left[-\exp\left\{-i\left(\frac{-\zeta(1-2y(1-\eta))}{y(1-\eta)}x_D
 -\frac{\eta(1-y)}{\left(1-\eta\right)}x_L
 +\left(\zeta-1\right)\frac{x_M}{1-y}
 -\zeta\frac{\eta y^2}{1-y}x_L\right) P^+\left({y'}_1^--{y}_1^-\right)
 \right\}\right] 
 \end{eqnarray}
 %
 % \checkmark %%%%%%%%%%%%%%%%%%%%%%%%%%%%%%%%%%%%%%%%%%%%%%%%%%%%%%%%%%%%%%%%%%%%%%%%%%%%%%%%%%%%%%%
 
 There are also three cuts in Fig.[\ref{base_13}], and the contributions are,

  \begin{figure}[ht]
   \begin{center}
 \includegraphics[width=8cm,height=4cm]{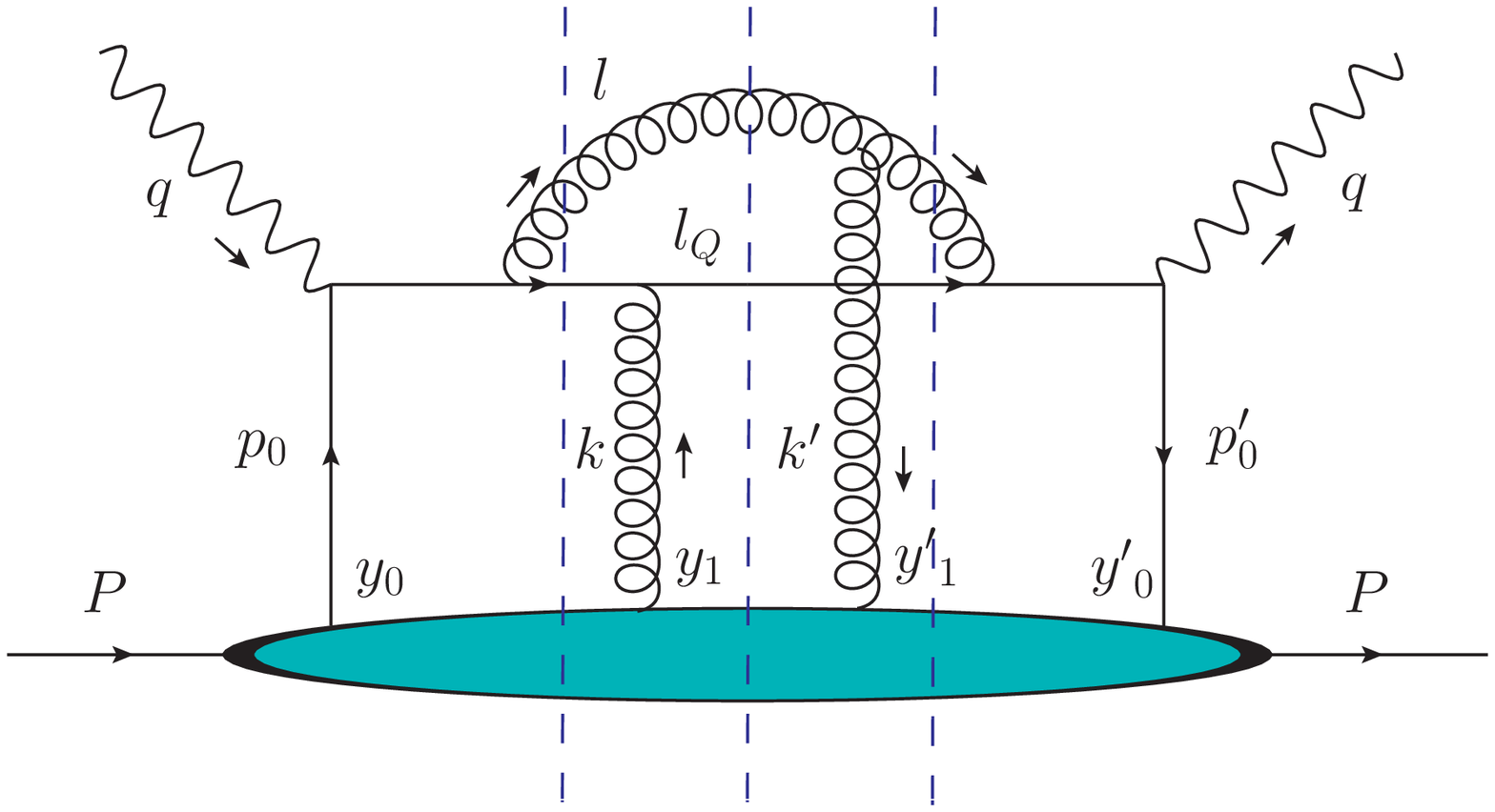}
 \caption{}
      \label{base_13}
  \end{center}
 \end{figure}
 \begin{eqnarray}
 {\cal C}_{13}^{c,l,r} &=& \frac{\alpha_s^2}{N_c}\left[-\frac{C_A}{2}\right]
                  \int dl_\perp^2 P(y) 
                  \frac{[l_\perp\left(l_\perp-k_\perp\right)+ \kappa y^4M^2]}
                  {[l_\perp^2+y^2M^2][\left(l_\perp-k_\perp\right)^2+y^2(1-\eta)^2M^2]} {\bar I}_{33}^{c,l,r}
 \end{eqnarray}
 \begin{eqnarray}
 {\bar I}_{13}^{c} &=& \exp\left[i\left(x_B+x_L+\frac{x_M}{1-y}\right)P^+\left({y'}_0^--{y}_0^-\right)\right]
 \exp\left[i\left(\zeta x_D+\left(\zeta-1\right)\frac{x_M}{1-y}
 -\zeta\frac{\eta y^2}{1-y}x_L\right)P^+\left({y'}_1^--{y}_1^-\right)\right] \theta\left({y}_0^--{y}_1^-\right)  \nn \\
 && \times ~ \theta\left({y'}_0^--{y'}_1^-\right) \left[1-\exp\left[-i\left(x_L+\frac{y}{1-y}x_M\right)P^+
 \left({y}_1^--{y}_0^-\right)\right]\right]  \nn \\
 && \left[\exp\left\{-i\left(\frac{\zeta}{y(1-\eta)}x_D
 +\frac{\eta(1-y)}{\left(1-\eta\right)}x_L
 +\left(\zeta-1\right)\frac{x_M}{1-y}
 -\zeta\frac{\eta y^2}{1-y}x_L\right) P^+\left({y'}_1^--{y'}_0^-\right)
 \right\}   \right. \nn \\
 && \left. -\exp\left[i\left(x_L+\frac{y}{1-y}x_M\right)P^+\left({y'}_1^--{y'}_0^-\right)\right]\right]   
 \end{eqnarray}
 \begin{eqnarray}
 {\bar I}_{13}^{l} &=& \exp\left[i\left(x_B+x_L+\frac{x_M}{1-y}\right)P^+\left({y'}_0^--{y}_0^-\right)\right]
 \exp\left[i\left(\zeta x_D+\left(\zeta-1\right)\frac{x_M}{1-y}
 -\zeta\frac{\eta y^2}{1-y}x_L\right)P^+\left({y'}_1^--{y}_1^-\right)\right] \theta\left({y'}_1^--{y}_1^-\right)  \nn \\
 && \times ~ \theta\left({y'}_0^--{y'}_1^-\right) (-1) 
  \left[\exp\left\{-i\left(\frac{\zeta}{y(1-\eta)}x_D
 +\frac{\eta(1-y)}{\left(1-\eta\right)}x_L
 +\left(\zeta-1\right)\frac{x_M}{1-y}
 -\zeta\frac{\eta y^2}{1-y}x_L\right) P^+\left({y'}_1^--{y'}_0^-\right)
 \right\}   \right. \nn \\
 && \left. -\exp\left[i\left(x_L+\frac{y}{1-y}x_M\right)P^+\left({y'}_1^--{y'}_0^-\right)\right]\right]  
 \end{eqnarray}
 \begin{eqnarray}
 {\bar I}_{13}^{r} &=& \exp\left[i\left(x_B+x_L+\frac{x_M}{1-y}\right)P^+\left({y'}_0^--{y}_0^-\right)\right]
 \exp\left[i\left(\zeta x_D+\left(\zeta-1\right)\frac{x_M}{1-y}
 -\zeta\frac{\eta y^2}{1-y}x_L\right)P^+\left({y'}_1^--{y}_1^-\right)\right] \theta\left({y}_0^--{y}_1^-\right)  \nn \\
 && \times ~ \theta\left({y'}_1^--{y}_1^-\right) \left[\exp\left\{i\left(\frac{\zeta}{y(1-\eta)}x_D
 +\frac{\eta(1-y)}{\left(1-\eta\right)}x_L
 +\left(\zeta-1\right)\frac{x_M}{1-y}
 -\zeta\frac{\eta y^2}{1-y}x_L\right)P^+\left(y_1^--y_0^-\right)
 \right\} \right. \nn \\
 && \left. -\exp\left[-i\left(x_L+\frac{y}{1-y}x_M\right)P^+
 \left({y}_1^--{y}_0^-\right)\right]\right] \left[-\exp\left\{-i\left( \frac{-\zeta(1-2y(1-\eta))}{y(1-\eta)}x_D
 -\frac{\eta(1-y)}{\left(1-\eta\right)}x_L
 +\left(\zeta-1\right)\frac{x_M}{1-y}
  \right. \right. \right. \nn \\
 && \left. \left. \left.  -\zeta\frac{\eta y^2}{1-y}x_L\right) P^+\left({y'}_1^--{y}_1^-\right)
 \right\}\right] 
 \end{eqnarray}

 The contributions from the three cuts in Fig.[\ref{base_31}] are,

 % \checkmark %%%%%%%%%%%%%%%%%%%%%%%%%%%%%%%%%%%%%%%%%%%%%%%%%%%%%%%%%%%%%%%%%%%%%%%%%%%%%%%%%%%%%%%
  \begin{figure}[ht]
   \begin{center}
 \includegraphics[width=8cm,height=4cm]{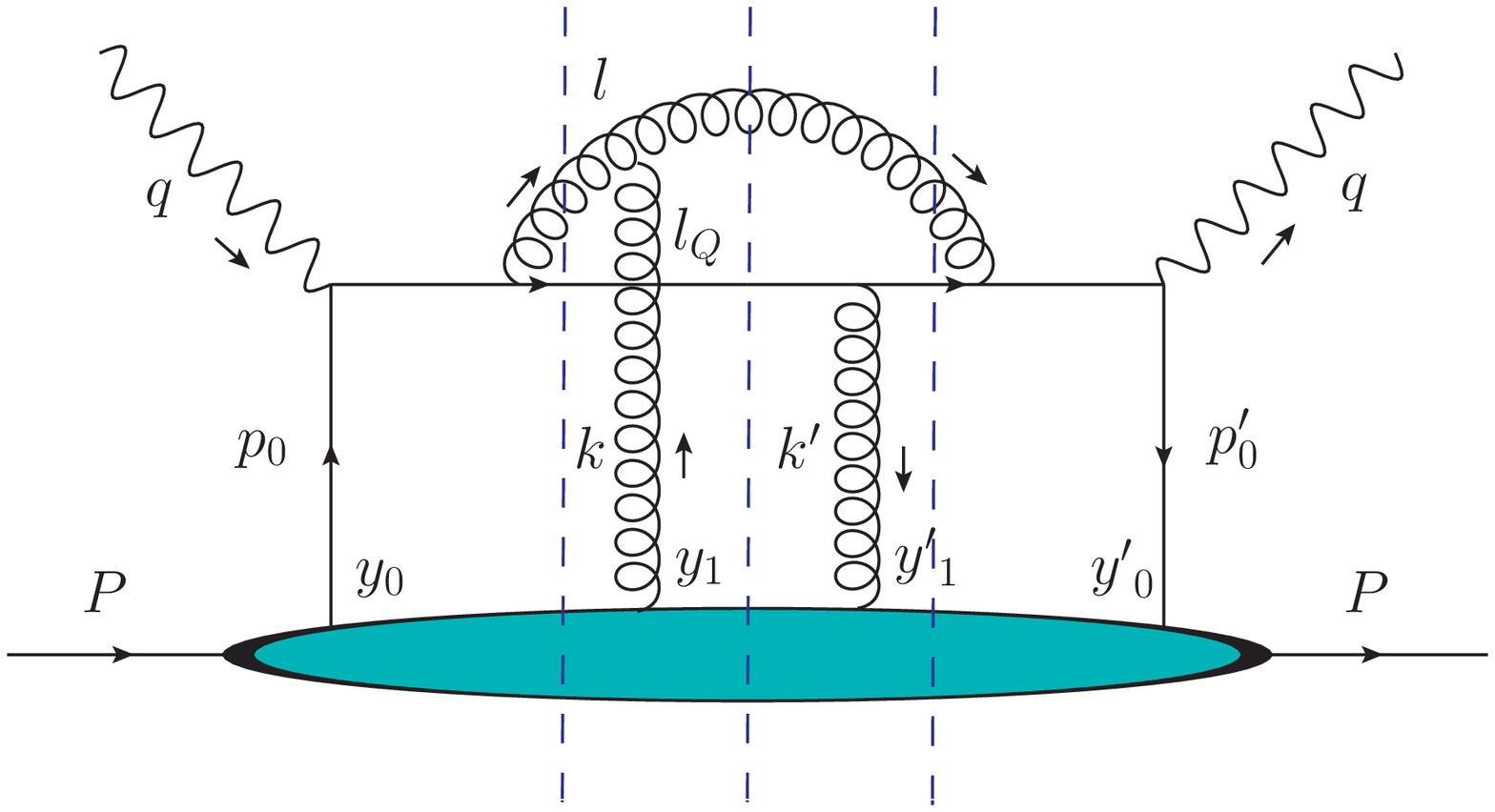}
 \caption{}
      \label{base_31}
  \end{center}
 \end{figure}
 \begin{eqnarray}
 {\cal C}_{31}^{c,l,r} &=& \frac{\alpha_s^2}{N_c}\left[-\frac{C_A}{2}\right]
                  \int dl_\perp^2 P(y) 
                   \frac{[\left(l_\perp-k_\perp\right)l_\perp+ \kappa y^4M^2]}
                   {[\left(l_\perp-k_\perp\right)^2+y^2(1-\eta)^2M^2][l_\perp^2+y^2M^2]} {\bar I}_{31}^{c,l,r} \nn       
 \end{eqnarray}
  \begin{eqnarray}
 {\bar I}_{31}^{c} &=& \exp\left[i\left(x_B+x_L+\frac{x_M}{1-y}\right)P^+\left({y'}_0^--{y}_0^-\right)\right]
 \exp\left[i\left(\zeta x_D+\left(\zeta-1\right)\frac{x_M}{1-y}
 -\zeta\frac{\eta y^2}{1-y}x_L\right)P^+\left({y'}_1^--{y}_1^-\right)\right] \theta\left({y}_0^--{y}_1^-\right)  \nn \\
 && \times ~ \theta\left({y'}_0^--{y'}_1^-\right) \left[\exp\left\{i\left(\frac{\zeta}{y(1-\eta)}x_D
 +\frac{\eta(1-y)}{\left(1-\eta\right)}x_L
 +\left(\zeta-1\right)\frac{x_M}{1-y}
 -\zeta\frac{\eta y^2}{1-y}x_L\right) P^+\left({y}_1^--{y}_0^-\right)
 \right\}   \right. \nn \\
 && \left. -\exp\left[-i\left(x_L+\frac{y}{1-y}x_M\right)P^+\left({y}_1^--{y}_0^-\right)\right]\right]  
 \left[1-\exp\left[i\left(x_L+\frac{y}{1-y}x_M\right)P^+
 \left({y'}_1^--{y'}_0^-\right)\right]\right] 
 \end{eqnarray}

 \begin{eqnarray}
  {\bar I}_{31}^{l} &=& \exp\left[i\left(x_B+x_L+\frac{x_M}{1-y}\right)P^+\left({y'}_0^--{y}_0^-\right)\right]
 \exp\left[i\left(\zeta x_D+\left(\zeta-1\right)\frac{x_M}{1-y}
 -\zeta\frac{\eta y^2}{1-y}x_L\right)P^+\left({y'}_1^--{y}_1^-\right)\right] \theta\left({y'}_0^--{y'}_1^-\right)  \nn \\
 && \times ~ \theta\left({y'}_1^--{y}_1^-\right)\left[-\exp\left\{-i\left(\frac{-\zeta(1-2y(1-\eta))}{y(1-\eta)}x_D
 -\frac{\eta(1-y)}{\left(1-\eta\right)}x_L
   +\left(\zeta-1\right)\frac{x_M}{1-y}
 -\zeta\frac{\eta y^2}{1-y}x_L\right)P^+\left({y'}_1^--{y}_1^-\right)
 \right\}\right]\nn \\
 &&  \left[\exp\left\{-i\left(\frac{\zeta}{y(1-\eta)}x_D
 +\frac{\eta(1-y)}{\left(1-\eta\right)}x_L
 +\left(\zeta-1\right)\frac{x_M}{1-y}
 -\zeta\frac{\eta y^2}{1-y}x_L\right)P^+\left({y'}_1^--{y'}_0^-\right)
 \right\} \right. \nn \\
 && \left. -\exp\left[i\left(x_L+\frac{y}{1-y}x_M\right)P^+
 \left({y'}_1^--{y'}_0^-\right)\right]\right]  
 \end{eqnarray}

  \begin{eqnarray}
 {\bar I}_{31}^{r} &=& \exp\left[i\left(x_B+x_L+\frac{x_M}{1-y}\right)P^+\left({y'}_0^--{y}_0^-\right)\right]
 \exp\left[i\left(\zeta x_D+\left(\zeta-1\right)\frac{x_M}{1-y}
 -\zeta\frac{\eta y^2}{1-y}x_L\right)P^+\left({y'}_1^--{y}_1^-\right)\right] \theta\left({y}_1^--{y'}_1^-\right)  \nn \\
 && \times ~ \theta\left({y}_0^--{y}_1^-\right)(-1)
  \left[\exp\left\{i\left(\frac{\zeta}{y(1-\eta)}x_D
 +\frac{\eta(1-y)}{\left(1-\eta\right)}x_L
 +\left(\zeta-1\right)\frac{x_M}{1-y}
 -\zeta\frac{\eta y^2}{1-y}x_L\right) P^+\left({y}_1^--{y}_0^-\right)
 \right\}   \right. \nn \\
 && \left. -\exp\left[-i\left(x_L+\frac{y}{1-y}x_M\right)P^+\left({y}_1^--{y}_0^-\right)\right]\right]  
 \end{eqnarray}

 % \checkmark %%%%%%%%%%%%%%%%%%%%%%%%%%%%%%%%%%%%%%%%%%%%%%%%%%%%%%%%%%%%%%%%%%%%%%%%%%%%%%%%%%%%%%%
  \begin{figure}[ht]
   \begin{center}
 \includegraphics[width=8cm,height=4cm]{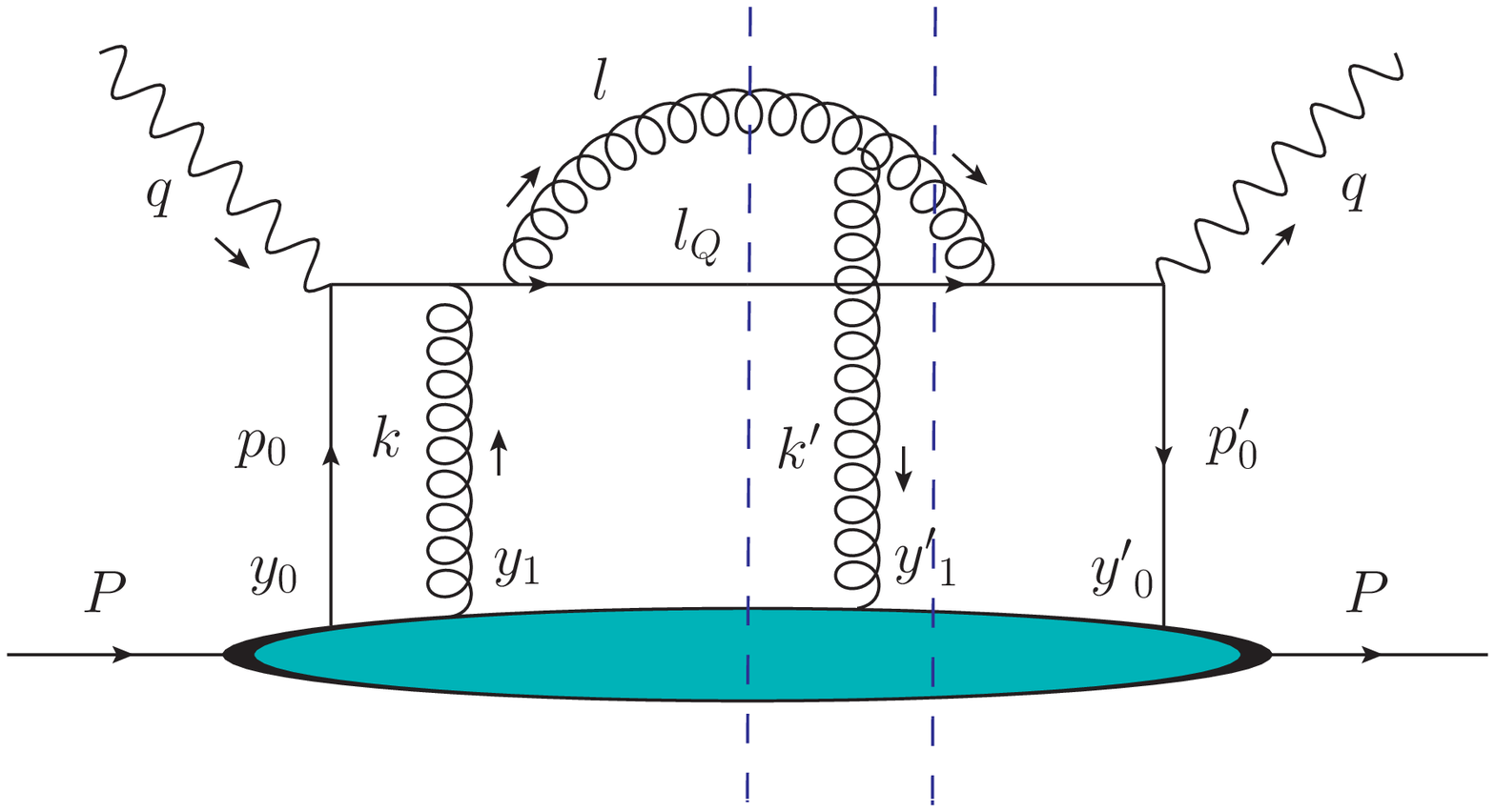}
 \caption{}
      \label{base_23}
  \end{center}
 \end{figure}

 Contributions from two cuts from Fig.[\ref{base_23}] are as follows,

 \begin{eqnarray}
 {\cal C}_{23}^{c} &=& \frac{\alpha_s^2}{N_c}\left[-\frac{C_A}{2}\right]
                  \int dl_\perp^2 \left(\frac{1+(1-y)^2}{y}\right) 
                  \frac{[\left(l_\perp-yk_\perp\right)\left(l_\perp-k_\perp\right)+ \kappa y^4M^2]}
                  {\left[\left(l_\perp-yk_\perp\right)^2+y^2M^2+2y\left(\frac{k^-}{l^-}\right)l_\perp^2\right]
                  [\left(l_\perp-k_\perp\right)^2+y^2(1-\eta)^2M^2]} {\bar I}_{23}^{c}  \nn 
 \end{eqnarray}
 
 \begin{eqnarray}
 {\bar I}_{23}^{c} &=& \exp\left[i\left(x_B+x_L+\frac{x_M}{1-y}\right)P^+\left({y'}_0^--{y}_0^-\right)\right]
 \exp\left[i\left(\zeta x_D+\left(\zeta-1\right)\frac{x_M}{1-y}
 -\zeta\frac{\eta y^2}{1-y}x_L\right)P^+\left({y'}_1^--{y}_1^-\right)\right]  \nn \\
 && \theta\left({y}_0^--{y}_1^-\right)  \theta\left({y'}_0^--{y'}_1^-\right) \left[-\exp\left[-i\left(x_L+\frac{y}{1-y}x_M\right)P^+
 \left({y}_1^--{y}_0^-\right)\right]\right]\nn \\
 && \left[\exp\left\{-i\left(\frac{\zeta}{y(1-\eta)}x_D
 +\frac{\eta(1-y)}{\left(1-\eta\right)}x_L
 +\left(\zeta-1\right)\frac{x_M}{1-y}
 -\zeta\frac{\eta y^2}{1-y}x_L\right) \right. \right. \nn \\
 && \left. \left. \times ~ P^+\left({y'}_1^--{y'}_0^-\right)
 \right\}-\exp\left[i\left(x_L+\frac{y}{1-y}x_M\right)P^+\left({y'}_1^--{y'}_0^-\right)\right]\right]    
 \end{eqnarray}

 \begin{eqnarray}
 {\cal C}_{23}^{r} &=& \frac{\alpha_s^2}{N_c}\left[-\frac{C_A}{2}\right]
                  \int dl_\perp^2 \left(\frac{1+(1-y)^2}{y}\right) 
                  \frac{[l_\perp\left(l_\perp-(1-y)k_\perp\right)+ \kappa y^4M^2]}
                  {\left[l_\perp^2+y^2M^2\right]
                  [\left(l_\perp-(1-y)k_\perp\right)^2+y^2(1-\eta)^2M^2]} {\bar I}_{23}^{r}  
 \end{eqnarray}
 
  \begin{eqnarray}
 {\bar I}_{23}^{r} &=& \exp\left[i\left(x_B+x_L+\frac{x_M}{1-y}\right)P^+\left({y'}_0^--{y}_0^-\right)\right]
 \exp\left[i\left(\zeta x_D+\left(\zeta-1\right)\frac{x_M}{1-y}
 -\zeta\frac{\eta y^2}{1-y}x_L\right)P^+\left({y'}_1^--{y}_1^-\right)\right] \nn \\
 && \times ~ \theta\left({y}_1^--{y'}_1^-\right) \theta\left({y}_0^--{y}_1^-\right) 
    \left[\exp\left[-i\left(x_L+\frac{y}{1-y}x_M\right)P^+
 \left({y'}_1^--{y}_0^-\right)-i\left(x_K-x_D\right)P^+\left({y}_1^--{y'}_1^-\right)\right] \right. \nn \\
 && \left. -\exp\left[-i\left(\frac{-\zeta(1-2y(1-\eta))}{y(1-\eta)}x_D
 -\frac{\eta(1-y)}{\left(1-\eta\right)}x_L\right)P^+\left({y'}_1^--{y}_1^-\right)-i\left(x_L+\frac{y}{1-y}x_M\right)P^+\left({y}_1^--{y}_0^-\right)\right]\right]   
 \end{eqnarray}

 % \checkmark %%%%%%%%%%%%%%%%%%%%%%%%%%%%%%%%%%%%%%%%%%%%%%%%%%%%%%%%%%%%%%%%%%%%%%%%%%%%%%%%%%%%%%%
  \begin{figure}[ht]
   \begin{center}
 \includegraphics[width=8cm,height=4cm]{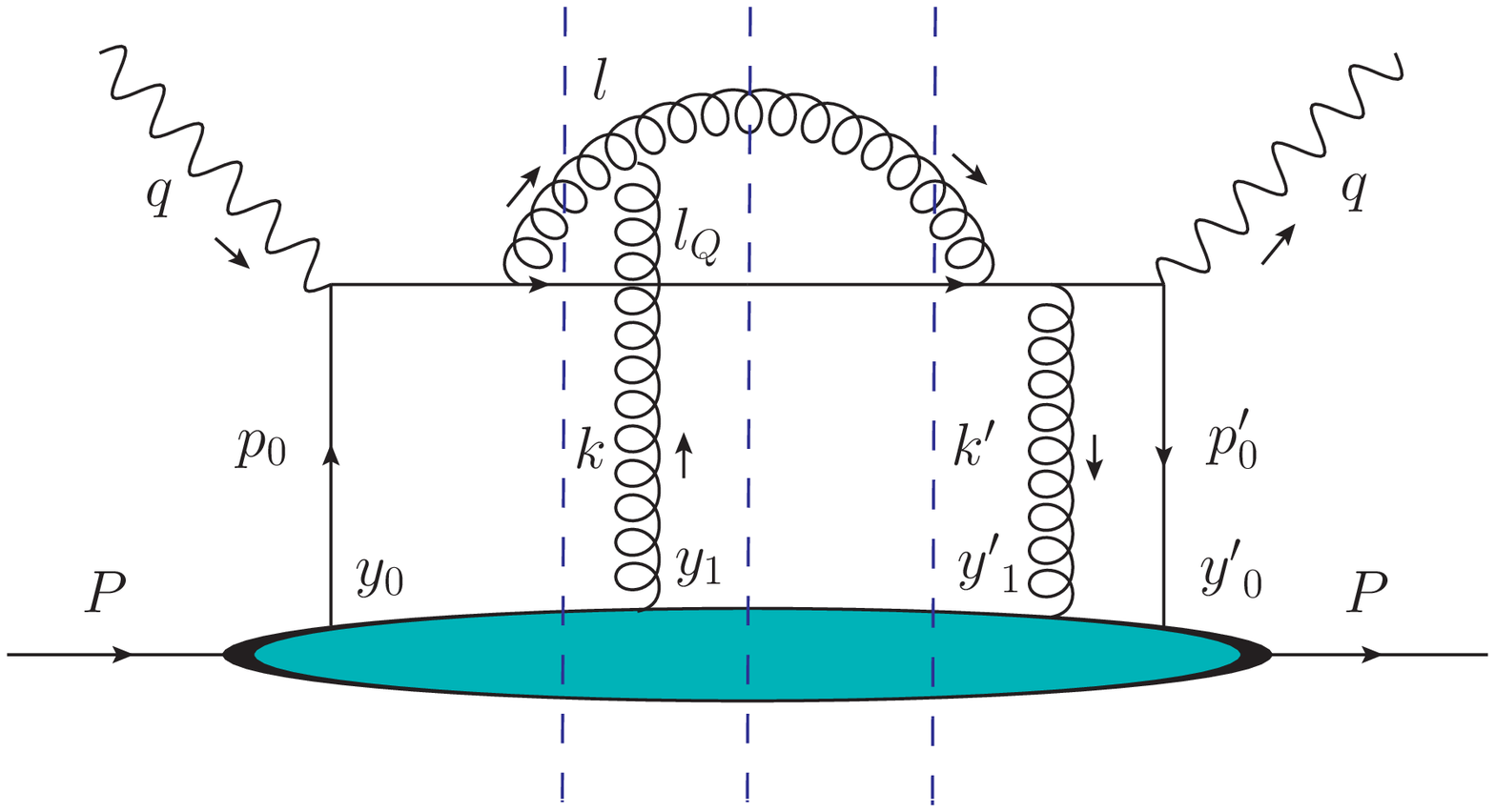}
 \caption{}
      \label{base_32}
  \end{center}
 \end{figure}

 Contributions of Fig.[\ref{base_32}] are,

 \begin{eqnarray}
 {\cal C}_{32}^{c} &=& \frac{\alpha_s^2}{N_c}\left[\frac{C_A}{2}\right]
                  \int dl_\perp^2 \left(\frac{1+(1-y)^2}{y}\right)
                  \frac{[\left(l_\perp-k_\perp\right)\left(l_\perp-yk_\perp\right)+ \kappa y^4M^2]}
                  {[\left(l_\perp-k_\perp\right)^2+y^2(1-k^-/l^-)^2M^2]
                  \left[\left(l_\perp-yk_\perp\right)^2+y^2M^2+2y\left(\frac{k^-}{l^-}\right)l_\perp^2\right]} 
                  {\bar I}_{32}^{c} \nn \\
 \end{eqnarray}
 
 \begin{eqnarray}
 {\bar I}_{32}^{c} &=& \exp\left[i\left(x_B+x_L+\frac{x_M}{1-y}\right)P^+\left({y'}_0^--{y}_0^-\right)\right]
 \exp\left[i\left(\zeta x_D+\left(\zeta-1\right)\frac{x_M}{1-y}
 -\zeta\frac{\eta y^2}{1-y}x_L\right)P^+\left({y'}_1^--{y}_1^-\right)\right] \theta\left({y}_0^--{y}_1^-\right)  \nn \\
 && \times ~ \theta\left({y'}_0^--{y'}_1^-\right) \left[\exp\left\{i\left(\frac{\zeta}{y(1-\eta)}x_D
 +\frac{\eta(1-y)}{\left(1-\eta\right)}x_L
 +\left(\zeta-1\right)\frac{x_M}{1-y}
 -\zeta\frac{\eta y^2}{1-y}x_L\right)P^+\left(y_1^--y_0^-\right)
 \right\} \right. \nn \\
 && \left. -\exp\left[-i\left(x_L+\frac{y}{1-y}x_M\right)P^+
 \left({y}_1^--{y}_0^-\right)\right]\right] \left[-\exp\left[i\left(x_L+\frac{y}{1-y}x_M\right)P^+\left({y'}_1^--{y'}_0^-\right)
 \right]\right]    
 \end{eqnarray}
 
  \begin{eqnarray}
 {\cal C}_{32}^{l} &=& \frac{\alpha_s^2}{N_c}\left[\frac{C_A}{2}\right]
                  \int dl_\perp^2 \left(\frac{1+(1-y)^2}{y}\right)
                  \frac{[\left(l_\perp-(1-y)k_\perp\right)l_\perp+ \kappa y^4M^2]}
                  {[\left(l_\perp-k_\perp\right)^2+y^2(1-\eta)^2M^2]
                  \left[\left(l_\perp-yk_\perp\right)^2+y^2M^2+2y\eta l_\perp^2\right]} 
                  {\bar I}_{32}^{l} \nn \\
 \end{eqnarray}
 
 \begin{eqnarray}
 {\bar I}_{12}^{l} &=& \exp\left[i\left(x_B+x_L+\frac{x_M}{1-y}\right)P^+\left({y'}_0^--{y}_0^-\right)\right]
 \exp\left[i\left(\zeta x_D+\left(\zeta-1\right)\frac{x_M}{1-y}
 -\zeta\frac{\eta y^2}{1-y}x_L\right)P^+\left({y'}_1^--{y}_1^-\right)\right] \nn \\
 && \times ~ \theta\left({y'}_1^--{y}_1^-\right) \theta\left({y'}_0^--{y'}_1^-\right) (-1)
    \left[\exp\left[i\left(x_L+\frac{y}{1-y}x_M\right)P^+
 \left({y}_1^--{y'}_0^-\right)+i\left(x_K-x_D\right)P^+\left({y'}_1^--y_1^-\right)\right] \right. \nn \\
 && \left. -\exp\left[i\left(\frac{-\zeta(1-2y(1-\eta))}{y(1-\eta)}x_D
 -\frac{\eta(1-y)}{\left(1-\eta\right)}x_L\right)P^+\left({y}_1^--{y'}_1^-\right)
 +i\left(x_L+\frac{y}{1-y}x_M\right)P^+\left({y'}_1^--{y'}_0^-\right)\right]\right]   
 \end{eqnarray}

 % \checkmark %%%%%%%%%%%%%%%%%%%%%%%%%%%%%%%%%%%%%%%%%%%%%%%%%%%%%%%%%%%%%%%%%%%%%%%%%%%%%%%%%%%%%%%
  \begin{figure}[ht]
   \begin{center}
 \includegraphics[width=8cm,height=4cm]{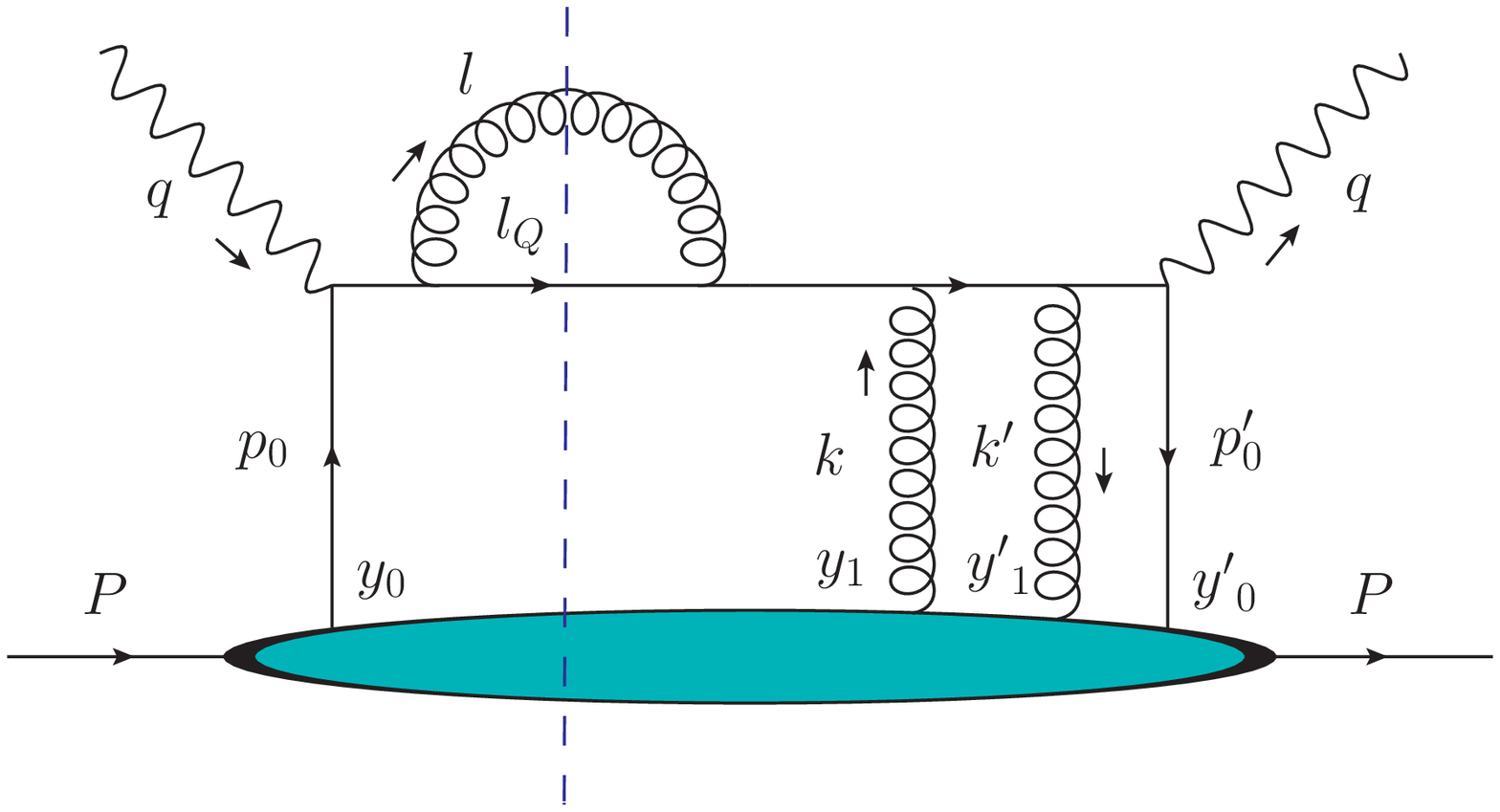}
 \caption{}
      \label{base_10}
  \end{center}
 \end{figure}

 Fig.[\ref{base_10}] has one possible cut,

  % \checkmark %%%%%%%%%%%%%%%%%%%%%%%%%%%%%%%%%%%%%%%%%%%%%%%%%%%%%%%%%%%%%%%%%%%%%%%%%%%%%%%%%%%%%%%
 \begin{eqnarray}
 {\cal C}_{10}^{l} &=& \frac{\alpha_s^2}{N_c}\left[C_F\right]
                   \int dl_\perp^2 P(y)
                   \frac{[l_\perp^2+ \kappa y^4M^2]}{(l_\perp^2+y^2M^2)^2} 
                   ~ {\bar I}_{10}^{l}
 \end{eqnarray}
 
  % \checkmark %%%%%%%%%%%%%%%%%%%%%%%%%%%%%%%%%%%%%%%%%%%%%%%%%%%%%%%%%%%%%%%%%%%%%%%%%%%%%%%%%%%%%%%
 \begin{eqnarray}
 {\bar I}_{10}^{l} &=& \exp\left[i\left(x_B+x_L+\frac{x_M}{1-y}\right)P^+\left({y'}_0^--{y}_0^-\right)\right]
 \exp\left[i\left(\zeta x_D+\left(\zeta-1\right)\frac{x_M}{1-y}
 -\zeta\frac{\eta y^2}{1-y}x_L\right)P^+\left({y'}_1^--{y}_1^-\right)\right] \theta\left({y'}_1^--{y}_1^-\right)  \nn \\
 && \times ~ \theta\left({y'}_0^--{y'}_1^-\right) 
    \left[-\exp\left[-i\left(x_K-x_D\right)P^+\left({y'}_1^--y_1^-\right)\right]\exp\left[i\left(x_L+\frac{y}{1-y}x_M\right)
    P^+\left({y}_1^--{y'}_0^-\right)\right]\right]
 \end{eqnarray}

 % \checkmark %%%%%%%%%%%%%%%%%%%%%%%%%%%%%%%%%%%%%%%%%%%%%%%%%%%%%%%%%%%%%%%%%%%%%%%%%%%%%%%%%%%%%%%
  \begin{figure}[ht]
   \begin{center}
 \includegraphics[width=8cm,height=4cm]{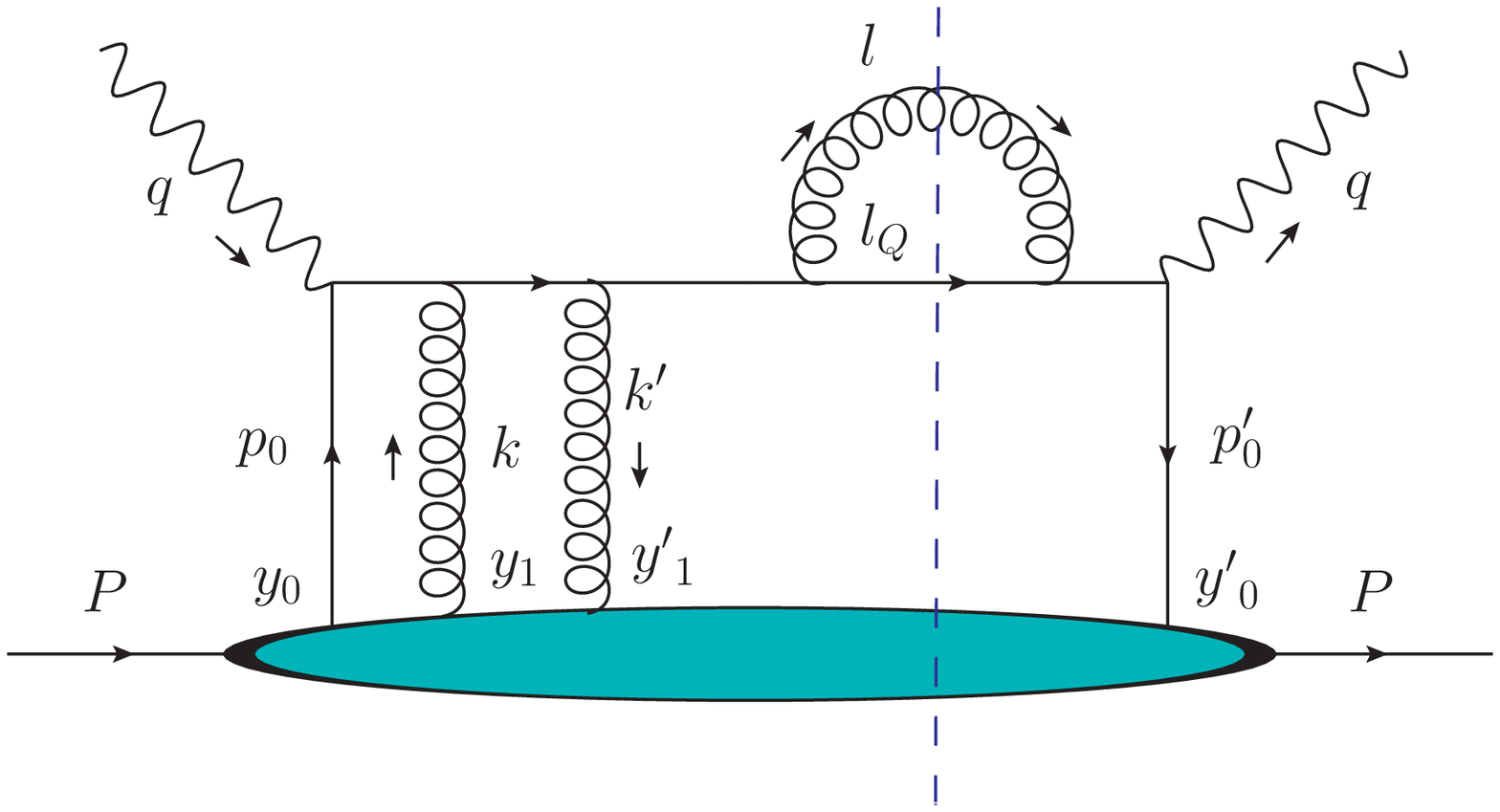}
 \caption{}
      \label{base_01}
  \end{center}
 \end{figure}

 Fig.[\ref{base_01}] also has single possible cut, 
 \begin{eqnarray}
 {\cal C}_{01}^{l} &=& \frac{\alpha_s^2}{N_c}\left[C_F\right]
                   \int dl_\perp^2 P(y)
                   \frac{[l_\perp^2+ \kappa y^4M^2]}{(l_\perp^2+y^2M^2)^2} 
                   ~ {\bar I}_{10}^{l}
 \end{eqnarray}
 
   % \checkmark %%%%%%%%%%%%%%%%%%%%%%%%%%%%%%%%%%%%%%%%%%%%%%%%%%%%%%%%%%%%%%%%%%%%%%%%%%%%%%%%%%%%%%%
 \begin{eqnarray}
 {\bar I}_{01}^{r} &=& \exp\left[i\left(x_B+x_L+\frac{x_M}{1-y}\right)P^+\left({y'}_0^--{y}_0^-\right)\right]
 \exp\left[i\left(\zeta x_D+\left(\zeta-1\right)\frac{x_M}{1-y}
 -\zeta\frac{\eta y^2}{1-y}x_L\right)P^+\left({y'}_1^--{y}_1^-\right)\right] \theta\left({y'}_1^--{y}_1^-\right)  \nn \\
 && \times ~ \theta\left({y'}_0^--{y'}_1^-\right) 
    \left[-\exp\left[i\left(x_K-x_D\right)P^+\left({y'}_1^--{y}_1^-\right)\right]\exp\left[-i\left(x_L+\frac{y}{1-y}x_M\right)P^+\left({y'}_1^--{y}_0^-\right)\right]\right]
 \end{eqnarray}
 \newpage

 \end{widetext}


\begin{thebibliography}{99}
 %%%%%%%%%%%%%%%%%%%%%%%%%%%%%%%%%%%%%%%%%%%%%%%%%%%%%%%%%%%%%%%%%%%
  
  
  %\cite{Wang:1991xy}
\bibitem{Wang:1991xy} 
  X.~N.~Wang and M.~Gyulassy,
  %``Gluon shadowing and jet quenching in A + A collisions at s**(1/2) = 200-GeV,''
  Phys.\ Rev.\ Lett.\  {\bf 68}, 1480 (1992).
  %%CITATION = PRLTA,68,1480;%%
  %657 citations counted in INSPIRE as of 31 août 2015
  
  %\cite{Gyulassy:1993hr}
\bibitem{Gyulassy:1993hr} 
  M.~Gyulassy and X.~n.~Wang,
  %``Multiple collisions and induced gluon Bremsstrahlung in QCD,''
  Nucl.\ Phys.\ B {\bf 420}, 583 (1994)
  [nucl-th/9306003].
  %%CITATION = NUCL-TH/9306003;%%
  %590 citations counted in INSPIRE as of 31 août 2015

%\cite{Baier:1996kr}
\bibitem{Baier:1996kr} 
  R.~Baier, Y.~L.~Dokshitzer, A.~H.~Mueller, S.~Peigne and D.~Schiff,
  %``Radiative energy loss of high-energy quarks and gluons in a finite volume quark - gluon plasma,''
  Nucl.\ Phys.\ B {\bf 483}, 291 (1997)
  [hep-ph/9607355].
  %%CITATION = HEP-PH/9607355;%%
  %653 citations counted in INSPIRE as of 31 août 2015

%\cite{Zakharov:1996fv}
\bibitem{Zakharov:1996fv} 
  B.~G.~Zakharov,
  %``Fully quantum treatment of the Landau-Pomeranchuk-Migdal effect in QED and QCD,''
  JETP Lett.\  {\bf 63}, 952 (1996)
  [hep-ph/9607440].
  %%CITATION = HEP-PH/9607440;%%
  %395 citations counted in INSPIRE as of 31 août 2015



  %\cite{Majumder:2010qh}
\bibitem{Majumder:2010qh} 
  A.~Majumder and M.~Van Leeuwen,
%  ``The Theory and Phenomenology of Perturbative QCD Based Jet Quenching,''
  Prog.\ Part.\ Nucl.\ Phys.\ A {\bf 66}, 41 (2011)
  [arXiv:1002.2206 [hep-ph]].
  %%CITATION = ARXIV:1002.2206;%%
  %143 citations counted in INSPIRE as of 24 Apr 2015

 
    

%\cite{Wiedemann:2009sh}
\bibitem{Wiedemann:2009sh} 
  U.~A.~Wiedemann,
%  ``Jet Quenching in Heavy Ion Collisions,''
  Landolt-Bornstein {\bf 23}, 521 (2010)
  [arXiv:0908.2306 [hep-ph]].
  %%CITATION = ARXIV:0908.2306;%%
  %77 citations counted in INSPIRE as of 24 Apr 2015


  %\cite{Armesto:2011ht}
\bibitem{Armesto:2011ht}
  N.~Armesto, B.~Cole, C.~Gale, W.~A.~Horowitz, P.~Jacobs, S.~Jeon, M.~van Leeuwen and A.~Majumder {\it et al.},
%  ``Comparison of Jet Quenching Formalisms for a Quark-Gluon Plasma 'Brick',''
  Phys.\ Rev.\ C {\bf 86} (2012) 064904
  [arXiv:1106.1106 [hep-ph]].
  %%CITATION = ARXIV:1106.1106;%%
  %78 citations counted in INSPIRE as of 24 Apr 2015
  
  
 %\cite{Qin:2009gw}
  \bibitem{Qin:2009gw} 
  G.~Y.~Qin and A.~Majumder,
%  ``A pQCD-based description of heavy and light flavor jet quenching,''
  Phys.\ Rev.\ Lett.\  {\bf 105}, 262301 (2010)
  [arXiv:0910.3016 [hep-ph]].
  %%CITATION = ARXIV:0910.3016;%%
  %26 citations counted in INSPIRE as of 03 Dec 2014
  
  
  %\cite{Djordjevic:2013pba}
  \bibitem{Djordjevic:2013pba} 
  M.~Djordjevic,
%  ``Heavy flavor puzzle at LHC: a serendipitous interplay of jet suppression and fragmentation,''
  Phys.\ Rev.\ Lett.\  {\bf 112}, no. 4, 042302 (2014)
  [arXiv:1307.4702 [nucl-th]].
  %%CITATION = ARXIV:1307.4702;%%
  %14 citations counted in INSPIRE as of 24 Apr 2015

  
  
  
  
  
  
  
  %\cite{Abelev:2006db}
\bibitem{Abelev:2006db} 
  B.~I.~Abelev {\it et al.}  [STAR Collaboration],
%  ``Erratum: Transverse momentum and centrality dependence of high-$p_T$ non-photonic electron suppression in Au+Au collisions at $\sqrt{s_{NN}} = 200$\,GeV,''
  Phys.\ Rev.\ Lett.\  {\bf 98}, 192301 (2007)
  [Phys.\ Rev.\ Lett.\  {\bf 106}, 159902 (2011)]
  [nucl-ex/0607012].
  %%CITATION = NUCL-EX/0607012;%%
  %450 citations counted in INSPIRE as of 24 Apr 2015
  
  %\cite{Adare:2006nq}
\bibitem{Adare:2006nq} 
  A.~Adare {\it et al.}  [PHENIX Collaboration],
%  ``Energy Loss and Flow of Heavy Quarks in Au+Au Collisions at s(NN)**(1/2) = 200-GeV,''
  Phys.\ Rev.\ Lett.\  {\bf 98}, 172301 (2007)
  [nucl-ex/0611018].
  %%CITATION = NUCL-EX/0611018;%%
  %549 citations counted in INSPIRE as of 24 Apr 2015

  
  
  
  %\cite{Djordjevic:2003zk} 
\bibitem{Djordjevic:2003zk} 
  M.~Djordjevic and M.~Gyulassy,
%  ``Heavy quark radiative energy loss in QCD matter,''
  Nucl.\ Phys.\ A {\bf 733}, 265 (2004)
  [nucl-th/0310076].
  %%CITATION = NUCL-TH/0310076;%%
  %176 citations counted in INSPIRE as of 24 Apr 2015

  %\cite{Djordjevic:2004nq}
\bibitem{Djordjevic:2004nq} 
  M.~Djordjevic, M.~Gyulassy and S.~Wicks,
%  ``The Charm and beauty of RHIC and LHC,''
  Phys.\ Rev.\ Lett.\  {\bf 94}, 112301 (2005)
  [hep-ph/0410372].
  %%CITATION = HEP-PH/0410372;%%
  %101 citations counted in INSPIRE as of 24 Apr 2015

  
  
  
  %\cite{ALICE:2012ab}
\bibitem{ALICE:2012ab} 
  B.~Abelev {\it et al.}  [ALICE Collaboration],
%  ``Suppression of high transverse momentum D mesons in central Pb-Pb collisions at $\sqrt{s_{NN}}=2.76$ TeV,''
  JHEP {\bf 1209}, 112 (2012)
  [arXiv:1203.2160 [nucl-ex]].
  %%CITATION = ARXIV:1203.2160;%%
  %218 citations counted in INSPIRE as of 24 Apr 2015

  
  
  
  
%\cite{Aamodt:2010jd}
\bibitem{Aamodt:2010jd} 
  K.~Aamodt {\it et al.}  [ALICE Collaboration],
%  ``Suppression of Charged Particle Production at Large Transverse Momentum in Central Pb--Pb Collisions at $\sqrt{s_{NN}} = 2.76$ TeV,''
  Phys.\ Lett.\ B {\bf 696}, 30 (2011)
  [arXiv:1012.1004 [nucl-ex]].
  %%CITATION = ARXIV:1012.1004;%%
  %402 citations counted in INSPIRE as of 24 Apr 2015
  
    
  %\cite{Mustafa:2004dr}
\bibitem{Mustafa:2004dr} 
  M.~G.~Mustafa,
%  ``Energy loss of charm quarks in the quark-gluon plasma: Collisional versus radiative,''
  Phys.\ Rev.\ C {\bf 72}, 014905 (2005)
  [hep-ph/0412402].
  %%CITATION = HEP-PH/0412402;%%
  %170 citations counted in INSPIRE as of 24 Apr 2015

  %\cite{Wicks:2005gt}
\bibitem{Wicks:2005gt} 
  S.~Wicks, W.~Horowitz, M.~Djordjevic and M.~Gyulassy,
%  ``Elastic, inelastic, and path length fluctuations in jet tomography,''
  Nucl.\ Phys.\ A {\bf 784}, 426 (2007)
  [nucl-th/0512076].
  %%CITATION = NUCL-TH/0512076;%%
  %393 citations counted in INSPIRE as of 24 Apr 2015

 
  
  %\cite{Abir:2011jb}
  \bibitem{Abir:2011jb} 
  R.~Abir, C.~Greiner, M.~Martinez, M.~G.~Mustafa and J.~Uphoff,
%  ``Soft gluon emission off a heavy quark revisited,''
  Phys.\ Rev.\ D {\bf 85}, 054012 (2012)
  [arXiv:1109.5539 [hep-ph]].
  %%CITATION = ARXIV:1109.5539;%%
  %24 citations counted in INSPIRE as of 03 Dec 2014

  %\cite{Abir:2012pu}
  \bibitem{Abir:2012pu} 
  R.~Abir, U.~Jamil, M.~G.~Mustafa and D.~K.~Srivastava,
%  ``Heavy quark energy loss and D-Mesons at RHIC and LHC energies,''
  Phys.\ Lett.\ B {\bf 715}, 183 (2012)
  [arXiv:1203.5221 [hep-ph]].
  %%CITATION = ARXIV:1203.5221;%%
  %18 citations counted in INSPIRE as of 03 Dec 2014
  
  %\cite{Zhang:2004qm}
  \bibitem{Zhang:2004qm} 
  B.~W.~Zhang, E.~k.~Wang and X.~N.~Wang,
%  ``Multiple parton scattering in nuclei: Heavy quark energy loss and modified fragmentation functions,''
  Nucl.\ Phys.\ A {\bf 757}, 493 (2005)
  [hep-ph/0412060].
  %%CITATION = HEP-PH/0412060;%%
  %23 citations counted in INSPIRE as of 03 Dec 2014
  
 
  
  
   %\cite{He:2012xz}
\bibitem{He:2012xz} 
  M.~He, R.~J.~Fries and R.~Rapp,
%  ``Non-perturbative Heavy-Flavor Transport at RHIC and LHC,''
  Nucl.\ Phys.\ A {\bf 910-911}, 409 (2013)
  [arXiv:1208.0256 [nucl-th]].
  %%CITATION = ARXIV:1208.0256;%%
  %16 citations counted in INSPIRE as of 24 Apr 2015

  
  
  %\cite{Cao:2011et}
\bibitem{Cao:2011et} 
  S.~Cao and S.~A.~Bass,
  %``Thermalization of charm quarks in infinite and finite QGP matter,''
  Phys.\ Rev.\ C {\bf 84}, 064902 (2011)
  [arXiv:1108.5101 [nucl-th]].
  %%CITATION = ARXIV:1108.5101;%%
  %28 citations counted in INSPIRE as of 31 août 2015
  
  
 %\cite{Moore:2004tg}
\bibitem{Moore:2004tg} 
  G.~D.~Moore and D.~Teaney,
  %``How much do heavy quarks thermalize in a heavy ion collision?,''
  Phys.\ Rev.\ C {\bf 71}, 064904 (2005)
  [hep-ph/0412346].
  %%CITATION = HEP-PH/0412346;%%
  %376 citations counted in INSPIRE as of 31 août 2015

%\cite{Baier:2002tc}
\bibitem{Baier:2002tc} 
  R.~Baier,
  %``Jet quenching,''
  Nucl.\ Phys.\ A {\bf 715}, 209 (2003)
  [hep-ph/0209038].
  %%CITATION = HEP-PH/0209038;%%
  %132 citations counted in INSPIRE as of 31 août 2015

%\cite{Majumder:2012sh}
\bibitem{Majumder:2012sh} 
  A.~Majumder,
  %``Calculating the jet quenching parameter ? in lattice gauge theory,''
  Phys.\ Rev.\ C {\bf 87}, 034905 (2013)
  [arXiv:1202.5295 [nucl-th]].
  %%CITATION = ARXIV:1202.5295;%%
  %35 citations counted in INSPIRE as of 31 août 2015


    %\cite{Majumder:2008zg}
\bibitem{Majumder:2008zg} 
  A.~Majumder,
  %``Elastic energy loss and longitudinal straggling of a hard jet,''
  Phys.\ Rev.\ C {\bf 80}, 031902 (2009)
  [arXiv:0810.4967 [nucl-th]].
  %%CITATION = ARXIV:0810.4967;%%
  %19 citations counted in INSPIRE as of 22 Dec 2014

  
  
  %\cite{Qin:2014mya}
\bibitem{Qin:2014mya} 
  G.~Y.~Qin and A.~Majumder,
%  ``Jet transport and photon bremsstrahlung via longitudinal and transverse scattering,''
  arXiv:1411.5642 [nucl-th].
  %%CITATION = ARXIV:1411.5642;%%
  %1 citations counted in INSPIRE as of 31 Mar 2015
  
  
  
  
  
  
  
 

  
  %\cite{Bauer:2000yr}
  \bibitem{Bauer:2000yr} 
  C.~W.~Bauer, S.~Fleming, D.~Pirjol and I.~W.~Stewart,
%  ``An Effective field theory for collinear and soft gluons: Heavy to light decays,''
  Phys.\ Rev.\ D {\bf 63}, 114020 (2001)
  [hep-ph/0011336].
  %%CITATION = HEP-PH/0011336;%%
  %813 citations counted in INSPIRE as of 19 Nov 2014
 
  %\cite{Bauer:2001ct}
  \bibitem{Bauer:2001ct} 
  C.~W.~Bauer and I.~W.~Stewart,
%  ``Invariant operators in collinear effective theory,''
  Phys.\ Lett.\ B {\bf 516}, 134 (2001)
  [hep-ph/0107001].
  %%CITATION = HEP-PH/0107001;%%
  %385 citations counted in INSPIRE as of 19 Nov 2014

  %\cite{Bauer:2001yt}
  \bibitem{Bauer:2001yt} 
  C.~W.~Bauer, D.~Pirjol and I.~W.~Stewart,
%  ``Soft collinear factorization in effective field theory,''
  Phys.\ Rev.\ D {\bf 65}, 054022 (2002)
  [hep-ph/0109045].
  %%CITATION = HEP-PH/0109045;%%
  %678 citations counted in INSPIRE as of 19 Nov 2014
  
  %\cite{Bauer:2002nz}
  \bibitem{Bauer:2002nz} 
  C.~W.~Bauer, S.~Fleming, D.~Pirjol, I.~Z.~Rothstein and I.~W.~Stewart,
%  ``Hard scattering factorization from effective field theory,''
  Phys.\ Rev.\ D {\bf 66}, 014017 (2002)
  [hep-ph/0202088].
  %%CITATION = HEP-PH/0202088;%%
  %264 citations counted in INSPIRE as of 19 Nov 2014
  
   %\cite{Abir:2014sxa}
  \bibitem{Abir:2014sxa} 
  R.~Abir, G.~D.~Kaur and A.~Majumder,
%  ``Multiple scattering of heavy-quarks in dense matter and the parametric prominence of drag,''
  arXiv:1407.1864 [nucl-th].
  %%CITATION = ARXIV:1407.1864;%%
  
  
  
 %\cite{Idilbi:2008vm}
\bibitem{Idilbi:2008vm} 
  A.~Idilbi and A.~Majumder,
  %``Extending Soft-Collinear-Effective-Theory to describe hard jets in dense QCD media,''
  Phys.\ Rev.\ D {\bf 80}, 054022 (2009)
  [arXiv:0808.1087 [hep-ph]].
  %%CITATION = ARXIV:0808.1087;%%
  %66 citations counted in INSPIRE as of 06 sept. 2015
  
  
  %\cite{D'Eramo:2010ak}
\bibitem{D'Eramo:2010ak} 
  F.~D'Eramo, H.~Liu and K.~Rajagopal,
  %``Transverse Momentum Broadening and the Jet Quenching Parameter, Redux,''
  Phys.\ Rev.\ D {\bf 84}, 065015 (2011)
  [arXiv:1006.1367 [hep-ph]].
  %%CITATION = ARXIV:1006.1367;%%
  %72 citations counted in INSPIRE as of 06 sept. 2015

  
  
  
  
  \bibitem{Qin:2012fua} 
  G.~-Y.~Qin and A.~Majumder,
  %``Parton Transport via Transverse and Longitudinal Scattering in Dense Media,''
  Phys.\ Rev.\ C {\bf 87}, 024909 (2013)
  [arXiv:1205.5741 [hep-ph]].
  %%CITATION = ARXIV:1205.5741;%%
  %8 citations counted in INSPIRE as of 19 Dec 2013
  

  
  
  
%   %\cite{Zhang:2003wk}
% \bibitem{Zhang:2003wk} 
% B.~-W.~Zhang, E.~Wang and X.~-N.~Wang,
%% ``Heavy quark energy loss in nuclear medium,''
% Phys.\ Rev.\ Lett.\  {\bf 93}, 072301 (2004)
% [nucl-th/0309040].
% %%CITATION = NUCL-TH/0309040;%%
% %111 citations counted in INSPIRE as of 08 Jul 2014
%

  
 
  
  %\cite{Dokshitzer:2001zm}
\bibitem{Dokshitzer:2001zm} 
  Y.~L.~Dokshitzer and D.~E.~Kharzeev,
  %``Heavy quark colorimetry of QCD matter,''
  Phys.\ Lett.\ B {\bf 519}, 199 (2001)
  [hep-ph/0106202].
  %%CITATION = HEP-PH/0106202;%%
  %568 citations counted in INSPIRE as of 08 Oct 2014
 
  %\cite{Wang:2001ifa}
\bibitem{Wang:2001ifa} 
  X.~N.~Wang and X.~f.~Guo,
  %``Multiple parton scattering in nuclei: Parton energy loss,''
  Nucl.\ Phys.\ A {\bf 696}, 788 (2001)
  [hep-ph/0102230].
  %%CITATION = HEP-PH/0102230;%%
  %360 citations counted in INSPIRE as of 23 Apr 2015
  

%\cite{Landau:1953um}
\bibitem{Landau:1953um} 
  L.~D.~Landau and I.~Pomeranchuk,
  %``Limits of applicability of the theory of bremsstrahlung electrons and pair production at high-energies,''
  Dokl.\ Akad.\ Nauk Ser.\ Fiz.\  {\bf 92}, 535 (1953).
  %%CITATION = DANKA,92,535;%%
  %580 citations counted in INSPIRE as of 04 sept. 2015

%\cite{Migdal:1956tc}
\bibitem{Migdal:1956tc} 
  A.~B.~Migdal,
%  ``Bremsstrahlung and pair production in condensed media at high-energies,''
  Phys.\ Rev.\  {\bf 103}, 1811 (1956).
  %%CITATION = PHRVA,103,1811;%%
  %576 citations counted in INSPIRE as of 24 Apr 2015


  %\cite{Majumder:2009ge}
\bibitem{Majumder:2009ge} 
  A.~Majumder,
%  ``Hard collinear gluon radiation and multiple scattering in a medium,''
  Phys.\ Rev.\ D {\bf 85}, 014023 (2012)
  [arXiv:0912.2987 [nucl-th]].
  %%CITATION = ARXIV:0912.2987;%%
  %25 citations counted in INSPIRE as of 05 Jan 2015
  
  


  
\end{thebibliography}
\end{document}